\documentclass[twocolumn]{aastex63}
\usepackage{url} 

\usepackage{txfonts}
\usepackage{threeparttable}
\usepackage{color,bm}

\newcommand{\eqref}[1]{(\ref{#1})}

\newcommand{\ko}[1]{\textcolor{black}{#1}}

\begin{document}

\title{Haze Formation on Triton}

\author[0000-0003-3290-6758]{Kazumasa Ohno}
\affil{Department of Astronomy and Astrophysics, University of California Santa Cruz, 1156 High St, Santa Cruz, CA 95064, USA}
\affil{Department of Earth and Planetary Sciences, Tokyo Institute of Technology, Meguro, Tokyo, 152-8551, Japan}

\author{Xi Zhang}
\affil{Department of Earth and Planetary Sciences, University of California Santa Cruz, 1156 High St, Santa Cruz, CA 95064, USA}

\author{Ryo Tazaki}
\affil{Anton Pannekoek Institute for Astronomy, University of Amsterdam, Science Park 904, 1098XH Amsterdam, The Netherlands}
\affil{Astronomical Institute, Tohoku University, 6-3, Aramaki, Aoba-ku, Sendai, Miyagi, 980-8578, Japan}

\author{Satoshi Okuzumi}
\affil{Department of Earth and Planetary Sciences, Tokyo Institute of Technology, Meguro, Tokyo, 152-8551, Japan}








\begin{abstract}
The largest moon of Neptune, Triton, possesses a cold and hazy atmosphere.
Since the discovery of near-surface haze layer during the Voyager fly in 1989, the haze formation mechanism has not been investigated in detail.
Here we provide the first haze microphysical model on Triton. 
Our model solves the evolution of both size and porosity distributions of haze particles in a self-consistent manner.
We simulated the formation of sphere and aggregate hazes with and without condensation of the C$_2$H$_4$ ice.
The haze particles can grow into fractal aggregates with mass-equivalent sphere sizes of ${\sim}0.1$--$1~{\rm \mu m}$ and fractal dimensions of $D_{\rm f}=1.8$--$2.2$.
The ice-free hazes cannot simultaneously explain both UV and visible observations of Voyager 2,
while including the condensation of C$_2$H$_4$ ices provides two better solutions. For ice aggregates, the required total haze mass flux is ${\sim}2\times{10}^{-15}~{\rm g~{cm}^{-2}~s^{-1}}$. For the icy sphere scenario, the column-integrated C$_2$H$_4$ production rate is ${\sim}8\times{10}^{-15}~{\rm g~{cm}^{-2}~s^{-1}}$, and the ice-free mass flux of ${\sim}6\times{10}^{-17}~{\rm g~{cm}^{-2}~s^{-1}}$.
The UV occultation observations at short wavelengths $<0.15~{\rm {\mu}m}$ may slightly favor the icy aggregates. 
Observations of the haze optical depth and the degree of forward scattering in UV and visible should be able to distinguish whether Triton's hazes are icy spheres or ice aggregates in future Triton missions.
\end{abstract}

\keywords{Planetary atmospheres (1244); Neptunian satellites (1098); Natural satellites (Solar system) (1089); Atmospheric clouds (2180)}

\section{Introduction} \label{sec:intro}

Organic aerosols (hereafter haze) produced via photochemistry of hydrocarbons are of great interest in studying atmospheric properties and surface environments.
The opacity of haze has crucial impacts on the radiative energy balance of atmospheres on Titan, Jupiter, and Pluto \citep[]{McKay+89_Titan,West+92,Zhang+15,Zhang+17}.
It has been suggested that the haze veiled Archean Earth \citep[e.g.,][]{Trainer+06,Zerkle+12} and played an important role in maintaining warm climates \citep[e.g.,][]{Sagan&Chyba97,Pavlov+01,Wolf&Toon10}.
In an atmospheric chemistry context, hazes act as loss sites of gaseous species via condensation \citep[e.g.,][]{Wong+17,Luspay-Kuti+17} and heterogeneous reactions \citep{Sekine+08a,Sekine+08b,Hong+18}.
Recent studies have also suggested that the presence of haze greatly impacts observations of exoplanetary atmospheres  \citep[e.g.,][]{Morley+15,Lavvas&Koskinen17,Kawashima&Ikoma18,Kawashima&Ikoma+19,Kawashima+19,Adams+19,Lavvas+19,Gao&Zhang20,Ohno&Kawashima20,Gao+20,Steinrueck+20,Lavvas&Arfaux21}.

In the outer solar system, observations of Titan, Pluto, and Triton provide important insights on organic haze formation in reduced (N$_2$-CH$_4$-CO) atmospheres.
The presence of haze on Titan was discovered by  ground-based observations \citep[e.g.,][]{Veverka73,Zellner73,Barker&Trafton73,Gillett+73,Danielson+73,Low&Rieke74,Caldwell75} and images from Voyager 1 \citep{Smith+81}.
Polarimetric and photometric observations of Titan's haze particles by Pioneer 11 and Voyager 1 \citep[][]{Rages&Pollack80,Tomasko&Smith82,Rages+83,West+83} are consistent with fractal aggregates---nonspherical particles constituted by numerous spherical monomers \citep[e.g.,][]{West&Smith91,Cabane+93,Rannou+95,Rannou+97,Karkoschka&Lorenz97}.
Cassini observations found that the Titan hazes extend from the ground to the ionosphere above 1100 km \citep[e.g.,][]{Tomasko+05,Liang+07,deKok+07,Vinatier+10}.
See \citet{McKay+01,Horst17} for more reviews on Titan's haze.

Pluto's hazes were discovered by the stellar occultations and have recently been investigated in detail by the New Horizons spacecraft \citep[e.g.,][]{Elliot+89,Elliot&Young92,Elliot+03,Gladstone+16,Cheng+17,Young+18}. 
The UV extinction coefficients of hazes are nearly proportional to the atmospheric (N$_2$) density from $26$ to $100~{\rm km}$ above the ground \citep{Young+18}.
The haze has a blue color that is consistent with Rayleigh scattering from particles with radii of ${\sim}0.01~{\rm \mu m}$ \citep{Gladstone+16}, whereas the strong forward scattering is consistent with particles with radii of ${\sim}0.5~{\rm {\mu}m}$ \citep{Cheng+17}.
This observational characteristic also indicates the aggregate nature of haze particles similar to the Titan haze \citep{Gladstone+16,Cheng+17}.

Triton has also been found to possess a near-surface haze layer in its thin atmosphere \citep{Yelle+95,Strobel&Summers95}.
The Voyager 2 imaging observations found that the optically thin hazes extend to an altitude of ${\sim}30~{\rm km}$ \citep{Smith+89}.
Using the high phase angle images, \citet{Pollack+90} estimated the particle size of ${\sim}0.1~{\rm \mu m}$, haze-scattering optical depth of ${\sim}0.003$, and the particle production rate of ${\sim}4.6\times{10}^{-15}~{\rm g~{cm}^{-2}~s^{-1}}$.
From disk-averaged photometry for a wavelength of $\lambda=0.414$--$0.561~{\rm \mu m}$, \citet{Hillier+90,Hillier+91} reported that Triton hazes have single a scattering albedo of nearly unity and cause strong forward scattering with an asymmetry factor of ${\sim}0.6$, although their results are highly influenced by discrete clouds near the ground \citep{Hillier&Veverka94}.
They also suggested that the scattering optical depth is nearly proportional to $\lambda^{-2}$.
From the disk-resolved photometry at the similar wavelength range, \citet{Rages&Pollack92} constrained the particle size of ${\sim}0.17~{\rm \mu m}$ and scattering optical depths of $0.001$--$0.01$ that is higher at shorter wavelengths.
Solar occultation observations at UV wavelengths ($\lambda=0.14$--$0.165~{\rm \mu m}$) constrained the extinction optical depth to ${\sim}0.024$, significantly higher than the scattering optical depth at visible wavelengths \citep{Herbert&Sandel91,Krasbopolsky+92}.
In sum, hazes on Titan, Pluto, and Triton all exhibit the wavelength-dependent opacity and the strong forward scattering.

Microphysical models have been used to investigate the haze formation processes and constrain fundamental parameters, such as the haze production rate and charge-to-radius ratio.
The models inferred the production rate of $0.5$--$3\times{10}^{-14}~{\rm g~{cm}^{-2}~s^{-1}}$ for Titan \citep[e.g.,][]{McKay+89_Titan,Toon+92,Rannou+97,Rannou+03,Lavvas+10} and $1.2\times{10}^{-14}~{\rm g~{cm}^{-2}~s^{-1}}$ for Pluto \citep{Gao+17}, respectively.
The microphysical models can also give insight on the degree of particle charge, which is associated with ionization processes in atmospheres \citep[e.g.,][]{Borucki+87,Borucki+06,Mishra+14}. 
Previous studies suggested the charge-to-radius ratio of $q_{\rm e}{\sim}15~{\rm e~{{\mu}m}^{-1}}$ for Titan \citep[e.g.,][]{Lavvas+10} and  $q_{\rm e}{\sim}30~{\rm e~{{\mu}m}^{-1}}$ for Pluto \citep[][]{Gao+17} to explain the degree of forward scattering of haze particles.

In contrast to Titan and Pluto, the Triton haze has not been thoroughly studied by a detailed microphysical model.
\citet{Strobel+90} performed photochemical calculations and estimated the haze production rate of ${\sim}4.7\times{10}^{-15}~{\rm g~{cm}^{-2}~s^{-1}}$ \citep[see also][]{Lyons+92,Krasnopolsky&Cruikshank95,Strobel&Summers95}, but they did not focus on the subsequent particle growth.
\citet{Krasbopolsky+92,Krasnopolsky93} calculated the condensation growth of settling haze particles.  
They assumed a refractive index of CH$_4$ for UV and C$_2$H$_4$ for the visible wavelength and suggested that the observed UV extinction coefficient and visible brightness coefficients could be explained if haze particles grow into $0.1$--$0.15~{\rm \mu m}$. 
However, they used the brightness coefficient derived from low phase angle photometry in \citet{Pollack+90}.
The high phase angle observation \citep{Rages&Pollack92}, which is more sensitive to haze properties, has not been compared with any haze microphysical models.
Moreover, they only considered spherical particles.
Triton hazes show scattering opacity increasing toward blue, implying small particles, and strong forward scattering, implying large particles. 
The coexistence of small and large particle properties may indicate the aggregate nature of Triton hazes.
However, to date, there has not been a study investigating whether Triton hazes are fractal aggregates.

\ko{Another interesting aspect of Triton hazes is that their properties are likely influenced by condensation of hydrocarbon ices.}
\citet{Sagan&Thompson84} pointed out that such ice condensation can occur in the lower atmosphere of Titan. 
\citet{Tomasko+08} found that the single-scattering albedo of Titan's haze increases with decreasing altitude between $140$ and $80~{\rm km}$, possibly due to ice condensation.
From electron conductivity of the Titan atmosphere, \citet{Borucki&Whitten08} suggested that the haze particles may be coated by condensed materials that cause less photoemission. 
In fact, infrared spectra identified several feature of organic ices in Titan's atmosphere \citep[e.g.,][]{Coustenis+99,deKok+14,Anderson+18}.
{  \citet{Lavvas+10,Lavvas+11a} demonstrated that the ice condensation is necessary to explain the spectral behavior of the haze optical depth on Titan below an altitude of $100~{\rm km}$.}
For Pluto's haze, the New Horizons observations suggested that the aggregates cannot explain the backscattering seen in the lower atmosphere, which may indicate that ice condensation alters the scattering properties of haze particles \citep{Cheng+17}.
\ko{Triton is a better site to study how ice condensation affects the haze formation, as the atmosphere is colder than Titan's and Pluto's atmospheres.}

In this study, we performed the first comprehensive investigations of haze formation on Triton using a detailed haze microphysical model. 
We developed a microphysical model that takes into account how the porosity and size distributions of the haze particles evolve in a self-consistent manner.
Furthermore, we quantified how the ice condensation affects haze vertical profiles and observational signatures. 
The organization of this paper is as follows.
We first review Triton's atmosphere and an adopted microphysical model in Section \ref{sec:method}.
We then present simulated haze vertical distributions without ice condensation and compare simulation results with the haze optical properties retrieved by observations in Section \ref{sec:result1}. 
As shown in Section \ref{sec:result1}, the observations cannot be explained by models without ice condensation.
We then propose two scenarios in Section \ref{sec:discussion} with ice condensation, ice spheres and ice aggregates, and demonstrate that the simulated ice particles could explain the Voyager observations. 
We discuss about caveats of this study and potential differences between Triton and Pluto hazes in Section \ref{sec:discussion2}. 
We finally summarize the key conclusions in Section \ref{sec:summary}.

\section{Method}\label{sec:method}
\subsection{Overview of Triton's Atmosphere and Haze Formation}\label{sec:method_intro}
\begin{figure}[t]
\centering
\includegraphics[clip, width=\hsize]{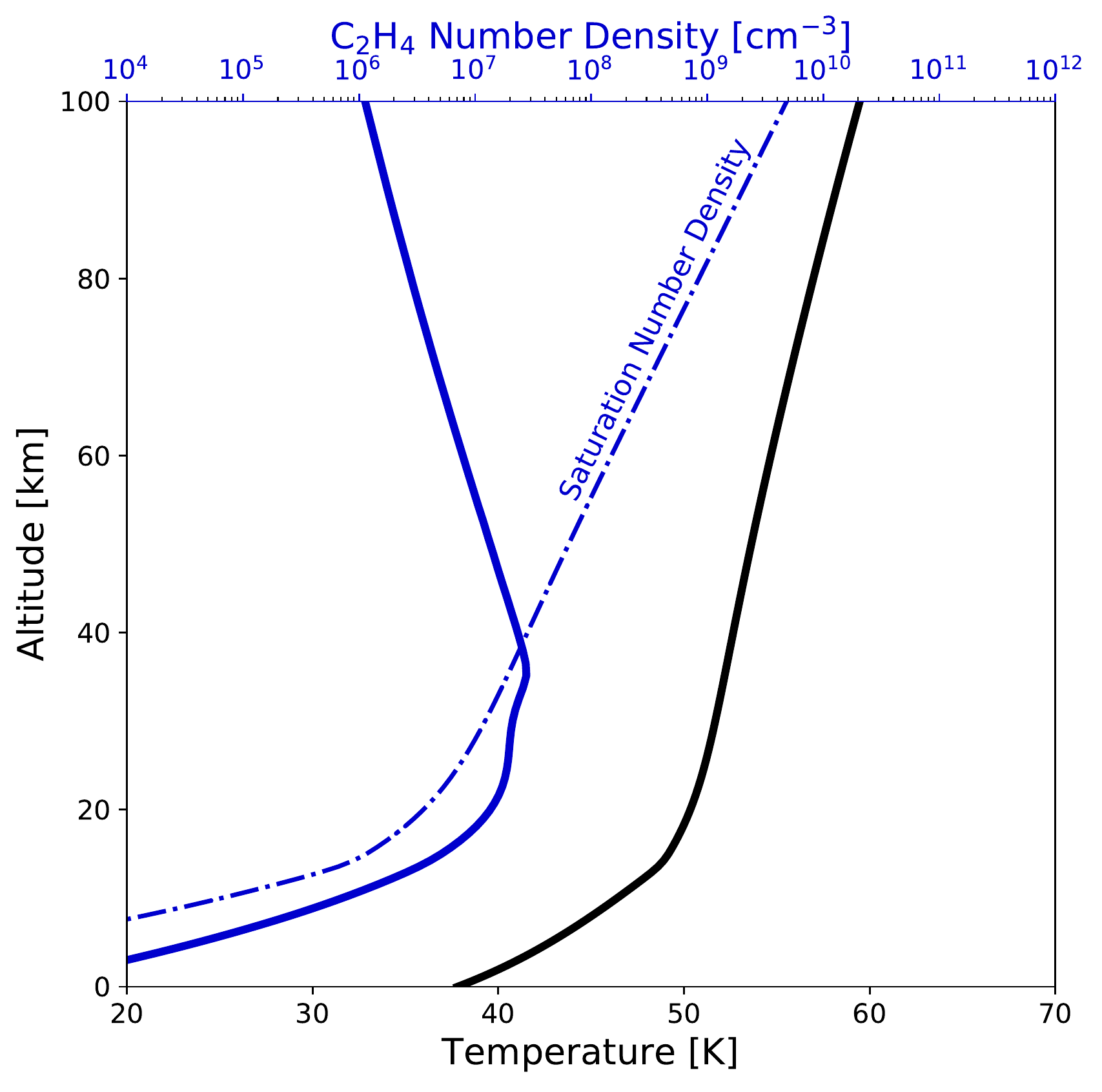}
\caption{Vertical atmospheric structure on Triton. The black line shows the atmospheric temperature from \citet{Strobel&Zhu17}. {  The solid blue line shows the number density of C$_2$H$_4$ molecules in our fiducial ice-ball simulations taken from Section \ref{sec:iceball}, where the column-integrated C$_2$H$_4$ production rate and ice-free mass flux are $F_{\rm vap}=3\times{10}^{-15}~{\rm g~{cm}^{-2}~s^{-1}}$ and $F_{\rm top}={10}^{-17}~{\rm g~{cm}^{-2}~s^{-1}}$, respectively.}
The dotted blue line shows the saturation number density of C$_2$H$_4$ calculated from the vapor pressure in \citet{Fray&Schmitt09_vapor-pressure}.}
\label{fig:PT_Triton}
\end{figure}

Triton's atmosphere is composed of about 99\% N$_2$, $0.01$\% CO, and $0.02$\% CH$_4$ \citep{Gladstone&Young19} and shows large seasonal variations \citep{Elliot+98,Elliot+00,Lellouch+10}.
In 1989 during the Voyager 2 flyby, the surface pressure and temperature were $1.8~{\rm Pa}$ and $38~{\rm K}$ in thermodynamical equilibrium with the surface ices \citep[e.g.,][]{Tyler+89,Broadfoot+89,Herbert&Sandel91}.
There are two kinds of atmospheric aerosols on Triton: discrete bright clouds at an altitude of $z<4~{\rm km}$ and optically thin haze extending to $z{\sim}30~{\rm km}$ \citep{Smith+89,Pollack+90}.
It is unclear whether the haze exists beyond $30~{\rm km}$ due to the detection limit of the Voyager 2 observations \citep{Pollack+90}.

The composition of the Triton's haze is under debate.
Previous studies usually assumed that the haze is composed of hydrocarbon ices  \citep[e.g.,][]{Pollack+90,Strobel+90,Krasbopolsky+92,Krasnopolsky&Cruikshank95}.
However, \citet{Rages&Pollack92} showed that photometric observations at high phase angles could be explained not only by conservatively scattering hazes but also by absorbing hazes.
The degeneracy stems from the fact that, for optically thin hazes, the scattered-light intensity is proportional to the scattering optical depth \citep[e.g.,][]{Liou02}.
This means that photometric observations cannot distinguish the scattering hazes with low extinction opacity from the absorbing hazes with high extinction opacity.

Haze formation can readily occur at the ionosphere according to Cassini observations on Titan \citep[][]{Liang+07,Lavvas+13}.
At the ionosphere, extreme-UV irradiation drives photochemistry and produces complex hydrocarbons and nitriles, which eventually form the initial haze particles.
Triton also undergoes the electron deposition from the magnetosphere of Neptune that can initiate auroral chemistry, enhancing the haze formation \citep{Thompson+89}.
The nucleated particles subsequently grow via surface chemistry until particle aggregation becomes efficient \citep{Lavvas+11}.
The predominant compositions of initial haze particles are highly uncertain, and previous studies suggested several candidates, such as polycyclic aromatic hydrocarbons and fullerenes \citep[e.g.,][]{Waite+07,Sittler+09}.
Triton has a dense ionosphere at an altitude of $\ga200~{\rm km}$ according to the Voyager Radio Science observations \citep{Tyler+89}. 
Such an upper atmosphere may be a birthplace of initial haze particles if the formation mechanism of Titan's haze also applies to Triton.

\ko{The formed haze particles may subsequently grow into fractal aggregates via mutual collisions and settle down to the ground.}
If this occurs, the shape of the aggregate is determined by the size ratios of collision pairs and trajectories of particle motions \citep[e.g.,][]{Meakin91,Cabane+93,Friedlander00,Okuzumi+09}.
For Titan hazes, the aggregates are constituted by monomers with radii of ${\sim}0.05~{\rm \mu m}$, and their sizes are ${\sim}0.1~{\rm \mu m}$ at an altitude below $150~{\rm km}$ \citep[e.g.,][]{Tomasko+05,Tomasko+08}.
For Pluto hazes, the New Horizons observations suggest the monomer sizes of ${\sim}0.01~{\rm {\mu}m}$ and aggregate sizes of $\ga0.1~{\rm \mu m}$ \citep{Gladstone+16,Cheng+17}.

The haze particles may further grow via condensation when the atmosphere is so cold that gaseous materials are saturated.
Triton's atmosphere is hotter at a higher altitude owing to the heating by the deposition of energetic particles from the magnetosphere of Neptune \citep[][]{Stevens+92,Strobel&Zhu17} (Figure \ref{fig:PT_Triton}).
However, ice condensation could readily occur in the cold lower atmosphere. 
Photochemical calculations suggest that condensation of hydrocarbons, such as $\rm C_2H_2$, $\rm C_2H_4$, and $\rm HCN$, can take place at $z< 60~{\rm km}$ of Triton's atmosphere \citep[e.g.,][]{Strobel+90}.
\ko{The most abundant condensable hydrocarbon is C$_2$H$_4$ \citep{Strobel&Summers95,Krasnopolsky&Cruikshank95} and it could condense to ice below $30~{\rm km}$} where it is supersaturated (see Figure \ref{fig:PT_Triton}).

\subsection{Overview of Adopted Haze Models}\label{sec:method_scenario}
\begin{table*}[t]
\begin{threeparttable}
  \caption{Haze formation scenarios in this study.}\label{table:1}
  \centering
  \begin{tabular}{l r r r r r} \hline
     & Particle Shape & Growth Process & Master Eqs.& Refractive index & Material density \\ \hline \hline
    Ice-free sphere& Compact sphere & Collision & \eqref{eq:basic} & Titan ``tholin" \tnote{a} & $1.00~{\rm g~{cm}^{-3}}$\\
    Ice-free aggregates& Fractal aggregates & Collision & \eqref{eq:basic}, \eqref{eq:basic2} & Titan ``tholin" \tnote{a} & $1.00~{\rm g~{cm}^{-3}}$\\
    Ice ball& Compact sphere & Collision, Condensation & \eqref{eq:basic_cond}, \eqref{eq:trans_vapor} & $(n,k)=(1.48,0)$\tnote{b} & $0.64~{\rm g~{cm}^{-3}}$\\
    Ice aggregates& Fractal aggregates & Collision &\eqref{eq:basic}, \eqref{eq:basic2} &
    $(n,k)=(1.48,0)$\tnote{b} & {  $0.64~{\rm g~{cm}^{-3}}$}\\ \hline
  \end{tabular}
      \begin{tablenotes}
        \raggedright
        \item[a] Refractive indices of the Titan ``tholin" are from \citet{Khare+84}.
        \item[b] Real refractive index corresponds to C$_2$H$_4$ ice measured at $\lambda=0.633~{\rm {\mu}m}$ \citep{Satorre+17}.
    \end{tablenotes}
 \end{threeparttable} 
\end{table*}

In this study, we investigate four possible scenarios of haze formation on Triton.
Each scenario and the adopted assumptions are summarized in Table \ref{table:1}.
We first examine the haze formation without ice condensation, as normally assumed for haze formation on Titan \citep[e.g.,][]{Toon+80,Toon+92,Cabane+92,Cabane+93,Rannou+03,Lavvas+10}.
Haze particles grow into either compact spheres or fractal aggregates via mutual collisions.
Such ice-free hazes might have particle refractive indices similar to Titan hazes.
Previous studies of haze formation on Pluto also adopted the ice-free haze model \citep{Rannou&Durry09,Gao+17,Rannou&West18}.
{  In this study, we adopt a particle material density of $1.0~{\rm g~{cm}^{-3}}$ and refractive index of Titan's tholin \citep{Khare+84} for the ice-free haze model following the previous studies.}

We also investigate the haze formation with the condensation of hydrocarbon ices.
We consider C$_2$H$_4$, the most abundant condensable hydrocarbon in Triton's lower atmosphere \citep[][]{Strobel+90,Strobel&Summers95,Krasnopolsky&Cruikshank95}.
In general, ice condensation {  can} affect both the optical properties and growth processes of haze particles. 
{To account for these effects, we consider the following two scenarios.}
In the first scenario, spherical hazes grow via condensation in addition to collisional growth. 
We refer to this model as an ``ice ball scenario''. 
In the second scenario, we consider ice condensation on the aggregates. We refer to this model as an ``ice aggregate scenario''. In fact, there are two possible outcomes of condensation on the aggregates: condensation reshapes the aggregates to the spheres or keeps the aggregate nature.
Even in the context of Earth's atmosphere, how condensation affects the morphology of aggregates is still an open question \citep[e.g.,][]{Ma+13,Heinson+17}.
In this study, we adopt a simplified model in which we assume that the condensation {only affects the optical properties of monomers and} does not affect the aggregate shape, such as their fractal dimensions. 
{In other words, ice condensation in our ice aggregate model only influences particle optical properties, not the particle growth directly \footnote{We note that the use of ice material density, different from the material density of ice-free hazes, can indirectly affect the particle growth.}. }
The most crucial difference between the ice aggregate case and the case of the ice-free aggregates is the particle optical properties; the ice aggregates likely have a much larger single-scattering albedo.
We leave detailed microphysical modeling of condensation on aggregates for future studies.
{For the icy haze models, we adopt the material density of C$_2$H$_4$ ice \citep[$0.64~{\rm g~{cm}^{-3}}$,][]{Satorre+17}.
Ice condensation also depends on the efficiency of heterogeneous nucleation onto haze particles, which we assume immediately occurs in saturated regions.
We will discuss the C$_2$H$_4$ nucleation properties in Section \ref{sec:nucleation}.
}

\subsection{Microphysical Model of Haze Collisional Growth}\label{sec:method1}
For all haze models examined in this study, we simulate the evolution of particle size distribution through collisional growth and gravitational settling.
The evolution of the size distribution is calculated by the { Smoluchowski} equation,
\begin{eqnarray}\label{eq:basic}
\nonumber
\frac{\partial n(m)}{\partial t}&=&\frac{1}{2}\int_{0}^{m}K(m',m-m')n(m')n(m-m')dm' \\
&&-n(m)\int_{0}^{\infty}K(m,m')n(m')dm'-\frac{\partial}{\partial z}\left[ v_{\rm t}n(m)\right]
\end{eqnarray}
where $n(m)dm$ is the number density of particles with masses between $m$ and $m+dm$, $K(m_{\rm 1},m_{\rm 2})$ is the collision kernel describing the collision rate between particles with masses of $m_{\rm 1}$ and $m_{\rm 2}$, and $v_{\rm t}$ is the terminal velocity of the particles.
The first and second terms stand for the gain and loss of haze particles via mutual collisions, and the third term expresses the gravitational settling.

{  
\begin{figure}[t]
\centering
\includegraphics[clip, width=\hsize]{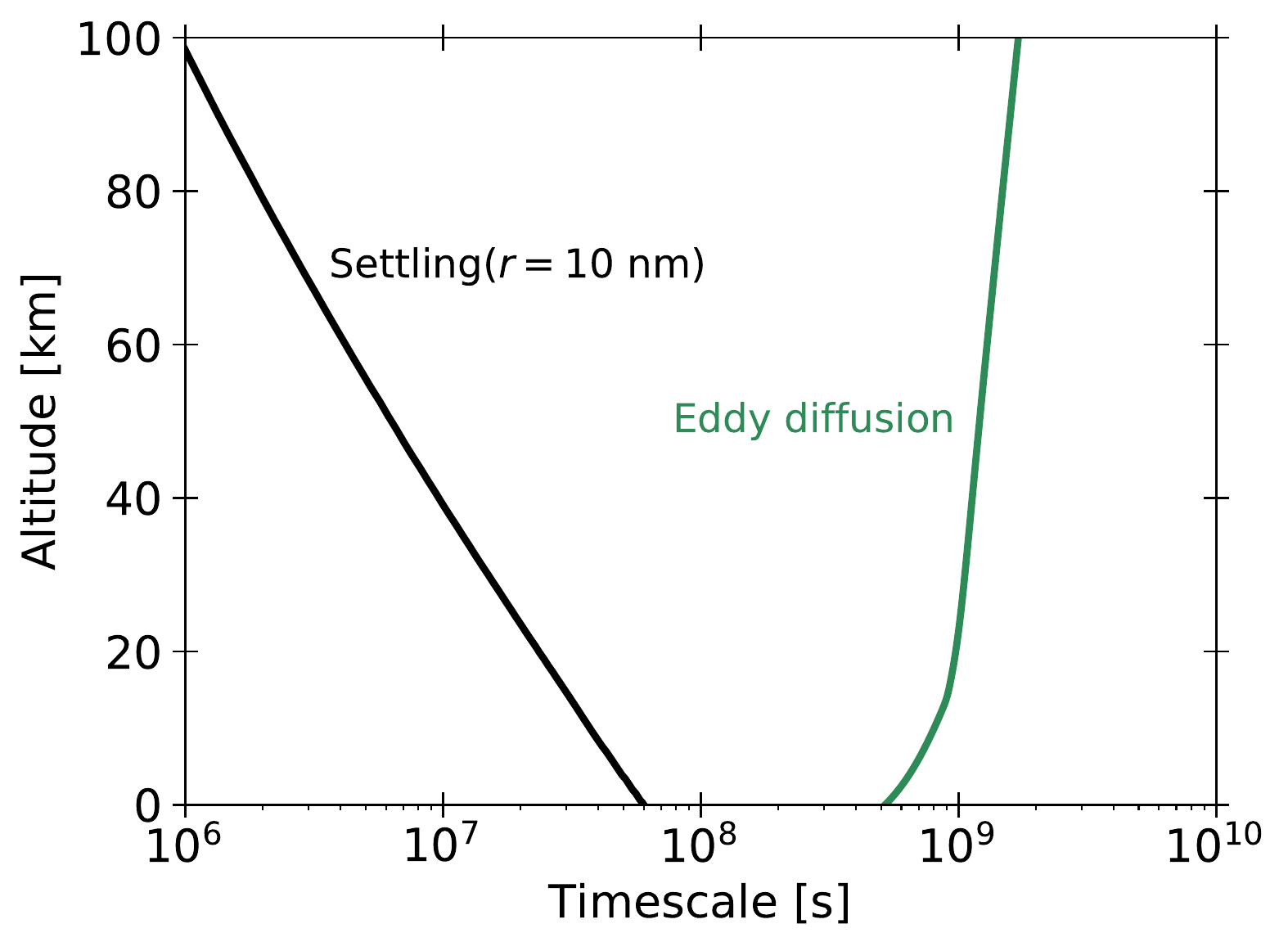}
\caption{
  Timescales of particle settling and eddy diffusion transport. The settling velocity is calculated by Equation \eqref{eq:vt}, where we set $r=10~{\rm nm}$ and $\rho_{\rm p}=1~{\rm g~{cm}^{-3}}$. The eddy diffusion coefficient is set to $K_{\rm z}=4\times{10}^{3}~{\rm {cm}^2~s^{-1}}$ \citep{Krasnopolsky&Cruikshank95}.
}
\label{fig:Timescale}
\end{figure}

We omit the diffusion transport of haze particles because the particle settling is much faster than the diffusional transport on Triton, similar to Pluto hazes \citep{Gao+17}.
Photochemical models have constrained the eddy diffusion coefficient on Triton to $K_{\rm z}=4\times{10}^{3}~{\rm {cm}^2~s^{-1}}$ using the observed CH$_4$ vertical profile \citep{Strobel+90,Krasnopolsky&Cruikshank95}.
Figure \ref{fig:Timescale} compares the particle settling timescale $H/v_{\rm t}$ with the eddy diffusion timescale $H^2/K_{\rm z}$, where $H$ is the pressure scale height.
The diffusion timescale is orders of magnitude longer than the settling timescale, even for tiny $10~{\rm nm}$ particles.
Thus, we can safely omit the diffusion transport for haze particles in our simulations.
}

We take into account the collisional growth of haze particles driven by the thermal Brownian motion for all simulations. 
The collision kernel is given by \citep{Seinfeld&Pandis06}
\begin{equation}\label{eq:K_coag}
    K(m_{\rm 1},m_{\rm 2})=\pi(r_{\rm 1}+r_{\rm 2})^2f_{\rm c}\sqrt{ \frac{8(m_{\rm 1}+m_{\rm 2})k_{\rm B}T}{\pi m_{\rm 1}m_{\rm 2}}},
\end{equation}
where $k_{\rm B}$ is the Boltzmann constant, $T$ is the temperature, and $f_{\rm c}$  is the correction factor accounting for the electrostatic repulsion introduced below.
We have assumed the free molecular flow regime. 
The assumption is valid for Triton's thin atmosphere unless particles grow into sizes of $\gg1~{\rm \mu m}$, much larger than the sizes of $\sim0.1~{\rm \mu m}$ suggested by photometric observations \citep{Rages&Pollack92}.
We also omit the collisional growth driven by the differential settling velocity, since it has negligible impacts for the parameter spaces examined in this study.

The particle charge regulates the collisional growth.
The importance of the particle charge was suggested for haze formation on Titan and Pluto \citep[e.g.,][]{Toon+80,Toon+92,Lavvas+10,Gao+17} and may also apply for Triton hazes, since there is a dense ionosphere \citep[][]{Tyler+89,Lyons+92}.
For the Brownian ballistic coagulation between particles with the same sign charge, the collision rate is reduced by\footnote{We note that the reduction factor of $f_{\rm c}=\Gamma/[\exp{(\Gamma)}-1]$ adopted in many previous studies was derived for diffusional coagulation \citep[see Section 18.2 of][]{pruppacher&Klett97} and not adequate for ballistic coagulation.} \citep[][]{Spitzer41}
\begin{equation}
    f_{\rm c}=\exp{(-\Gamma)},
\end{equation}
where $\Gamma$ is the ratio of the electrostatic repulsion energy to the collision energy, given by 
\begin{equation}\label{eq:Gamma_B}
    \Gamma = \frac{q_{\rm e}^2r_{\rm 1}r_{\rm 2}}{(r_{\rm 1}+r_{\rm 2})k_{\rm B}T},
\end{equation}
where $q_{\rm e}$ is the charge density defined as the ratio of the average electron numbers to a particle radius.
Previous studies suggested a charge density of $q_{\rm e}=15$--$30~{\rm e~{\mu m}^{-1}}$ to explain the particle sizes of the Titan and Pluto hazes \citep[e.g.,][]{Toon+92,Lavvas+10,Gao+17}.
The charge density is associated with the ionization and recombination processes of plasma particles in an atmosphere \citep[][]{Borucki+87,Borucki+06,Borucki&Whitten08,Mishra+14}.
In this study, we vary $q_{\rm e}$ as a free parameter for sensitivity studies.

In general, the terminal velocity depends on particle size, shape, gas density, and terminal velocity itself.
In Triton's tenuous atmosphere, where the mean free path of gas particles ($\sim750~{\rm \mu m}$ at the ground) is much larger than the expected sizes of aerosol particles, the terminal velocity can be approximated by \citep[e.g.,][]{Gao+17} 
\begin{equation}\label{eq:vt}
    v_{\rm t}\approx0.74\frac{\rho_{\rm p}gr}{\rho_{\rm g}C_{\rm s}}\frac{\pi r^2}{A},
\end{equation}
where $\rho_{\rm p}$ is the particle density, $g$ is the surface gravity, $\rho_{\rm g}$ is the gas density, $C_{\rm s}=\sqrt{8k_{\rm B}T/\pi m_{\rm g}}$ is the mean thermal velocity, $T$ is the temperature, $m_{\rm g}$ is the molecular mass, $r$ is the particle radius, and $A$ is the projected area of a particle.
\ko{The spheres have the projected area of $A=\pi r^2$.
We introduce how to evaluate $A$ for the aggregates in next section.}
For the ice-free spheres, we assume the particle density of $1~{\rm g~{cm}^{-3}}$ following previous studies \citep{Toon+80,Toon+92,Lavvas+10,Gao+17}.

\subsection{Porosity Evolution with Size Distribution}\label{method:volume_average}
Triton hazes may be aggregates constituted by numerous monomers similar to Titan hazes.
The number of monomers is defined by
\begin{equation}\label{eq:Df}
    N_{\rm mon}=k_{\rm 0}\left(\frac{r_{\rm agg}}{r_{\rm mon}} \right)^{D_{\rm f}},
\end{equation}
where $k_{\rm 0}{\sim}1$ is the prefactor, $r_{\rm mon}$ is the monomer radius, $r_{\rm agg}$ is the characteristic radius of an aggregate, and $D_{\rm f}$ is the fractal dimension.
Here $D_{\rm f}$ characterizes the shape of an aggregate; for example, a chain-like aggregate is characterized by $D_{\rm f}\approx1$, and a spherical aggregate is characterized by $D_{\rm f}\approx3$.
The actual value of $D_{\rm f}$ depends on the size ratio of collision pairs and trajectory of particle motions; for example, the aggregate formed via ballistic similar-sized collisions has $D_{\rm f}=1.7$--$2.2$ (a so-called cluster-cluster aggregate), while ballistic aggregate-monomer collisions yield spherical aggregates with $D_{\rm f}\approx3$ \citep[e.g.,][]{Meakin91,Cabane+93,Friedlander00,Okuzumi+09}.
Photometric observations suggested that the fractal dimensions of Titan hazes are close to $D_{\rm f}=2$ \citep{Rannou+97}.

In this study, we explicitly simulate the evolution of particle porosity instead of assuming a constant $D_{\rm f}$. 
To this end, we adopt the volume-averaging method proposed by \citet{Okuzumi+09}.
This method simulates the evolution of the mean volume of aggregates in each mass grid, $\overline{V}(m)$, by solving an additional { Smoluchowski} equation, given by
\begin{eqnarray}\label{eq:basic2}
\nonumber
\frac{\partial \overline{V}(m)n(m)}{\partial t}&=&\frac{1}{2}\int_{0}^{m}[V_{\rm 1+2}K](m',m-m')n(m')n(m-m')dm' \\
\nonumber
&&-\overline{V}(m)n(m)\int_{0}^{\infty}K(m,m')n(m')dm'\\
&&-\frac{\partial}{\partial z}\left[v_{\rm t}\overline{V}(m)n(m)\right],
\end{eqnarray}
where $V_{\rm 1+2}$ is the volume of an aggregate produced by the collision between aggregates with volumes $V_{\rm 1}$ and $V_{\rm 2}$.
The volume of an aggregate is defined by 
\begin{equation}\label{eq:V}
    \overline{V}=\frac{4}{3}\pi r_{\rm agg}^3.
\end{equation}
Since a newly formed aggregate contains voids within the bodies, the volume of the merged aggregate is described by
\begin{equation}\label{eq:V12}
    V_{\rm 1+2}=V_{\rm 1}+V_{\rm 2}+V_{\rm void},
\end{equation}
where $V_{\rm void}$ is the volume of newly formed voids within the aggregate.
For low-energy ballistic collisions, \citet{Okuzumi+09} derived an empirical relation of the void volume from direct $N$-body simulations of aggregate sticking, given by
\begin{equation}\label{eq:V_void}
    V_{\rm void}=\min{\left[0.99-1.03\ln{\left(\frac{2}{V_{\rm 1}/V_{\rm 2}+1}\right)}, 6.94 \right]}V_{\rm 2},
\end{equation}
where $V_{\rm 1}\geq V_{\rm 2}$.
In the limit of a similar-sized collision with $V_{\rm 1}=V_{\rm 2}$, Equation~\eqref{eq:V_void} yields $V_{\rm void}\approx V_{\rm 1}$ and thus $V_{\rm 1+2}=2.99V_{\rm 1}$.
This is consistent with the fractal relation of $V_{\rm 1+2}= (2m_{\rm 1}/m_{\rm 1})^{3/D_{\rm f}}V_{\rm 1}$ for $D_{\rm f}\approx1.9$, a typical fractal dimension resulting from ballistic cluster-cluster aggregations.
In the opposite limit of the large size ratio, $V_{\rm 2}/V_{\rm 1}\ll1$, Equation~\eqref{eq:V_void} yields $V_{\rm void}\approx 6.94V_{\rm 2}$. 
This leads to the formation of ballistic particle-cluster aggregates with volume filling factor of $\approx0.126$, as seen in $N$-body simulations \citep{Okuzumi+09}.
The volume-averaging method is based on the assumption that the volume distribution of aggregates is narrowly peaked at the mean volume in each mass grid.
This method well explains the fractal dimension of aggregates observed in full $N$-body simulations and experiments \citep{Okuzumi+09} and later applied to simulations of dust growth in protoplanetary disks \citep{Okuzumi+11,Okuzumi+12,Homma&Nakamoto18}. 
The aerosol science for the Earth's atmosphere also adopted a similar method to simulate the morphological evolution of soot aggregates \citep[e.g.,][]{Kostoglow+06}.
\ko{In this study, we ignore the restructure of the aggregates, as it only occurs for aggregates larger than $\sim{10}~{\rm {\mu}m}$ \citep{Ohno+20}.}

The microphysical terms presented in Section \ref{sec:method1} can apply to aggregates using the characteristic radius $r_{\rm agg}$ instead of a sphere radius \citep[][]{Cabane+93}.
The terminal velocity of an aggregate can also be approximated by Equation~\eqref{eq:vt} {  with the particle density evaluated as $\rho_{\rm p}=m/\overline{V}$}.
However, one should carefully evaluate the projected area of an aggregate $A$. 
This is because, for $D_{\rm f}<2$, the aggregates settle more slowly, as they are larger if one naively assumes $A=\pi r_{\rm agg}^2$. 
\citet{Okuzumi+09} provided an approximated formula of the projected area for general aggregates, given by
\begin{equation}\label{eq:A_project}
    A = \left( \frac{1}{\overline{A}_{\rm BCCA}} + \frac{1}{\pi r_{\rm agg}^2} - \frac{1}{\pi r_{\rm mon}^2N_{\rm mon}^{2/D_{\rm f,BCCA}}} \right)^{-1},
\end{equation}
where $\overline{A}_{\rm BCCA}$ and $D_{\rm f,BCCA}\approx1.9$ are the averaged projected area and fractal dimension of ballistic cluster-cluster aggregates.
The averaged area of the ballistic cluster-cluster aggregate can be well approximated by \citep{Minato+06}
\begin{equation}
\label{eq:sigma_sca}
    \frac{\overline{A}_{\rm BCCA}}{\pi r_{\rm mon}^2 N_{\rm mon}}=  
\left\{
\begin{array}{ll}
      12.5N_{\rm mon}^{-0.315}\exp{(-2.53N_{\rm mon}^{-0.0920})} & \text{($N_{\rm mon}<16$)} \\
      0.352 + 0.566N_{\rm mon}^{-0.138} & \text{($N_{\rm mon}\geq16$)}
    \end{array}
\right.
\end{equation}
Equation \eqref{eq:A_project} reduces to $A=A_{\rm BCCA}$ in the limit of the ballistic cluster-cluster aggregate (i.e., $D_{\rm f}\approx1.9$) and $A=\pi r_{\rm agg}^2$ in the limit of $D_{\rm f}\gg2$ and large $N_{\rm mon}$.
\citet{Suyama+12} verified the validity of the projected area formula using direct $N$-body simulations of sequential collisions of aggregates.

\subsection{Growth via Condensation}\label{sec:method_cond}
We additionally include condensation growth in the ice ball model (Table \ref{table:1}).
The condensation is described as an advection term in a mass space \citep[][]{Seinfeld&Pandis06,Lavvas+11a}, and
Equation \eqref{eq:basic} is rewritten as
\begin{eqnarray}\label{eq:basic_cond}
\nonumber
\frac{\partial n(m)}{\partial t}&=&\frac{1}{2}\int_{0}^{m}K(m',m-m')n(m')n(m-m')dm' \\
\nonumber
&&-n(m)\int_{\rm 0}^{\rm \infty}K(m,m')n(m')dm'-\frac{\partial}{\partial z}\left[ v_{\rm t}n(m)\right]\\
&&-\frac{\partial}{\partial m}\left[ \left( \frac{dm}{dt}\right)_{\rm cond}n(m)\right].
\end{eqnarray}
The last term stands for the condensation with a growth rate of $(dm/dt)_{\rm cond}$.
In the Triton's tenuous atmosphere, the condensation rate is described by the kinetic regime given by \citep[][]{Seinfeld&Pandis06}
\begin{equation}\label{eq:dmdt_cond}
    \left(\frac{dm}{dt}\right)_{\rm cond}=4\pi r^2 v_{\rm rel} \rho_{\rm s}\left[S-\exp{\left( \frac{2m_{\rm v}\sigma}{\rho_{\rm 0}k_{\rm B}Tr}\right)} \right],
\end{equation}
where $v_{\rm rel}=\sqrt{k_{\rm B}T/2\pi m_{\rm v}}$ is the mean relative velocity, $m_{\rm v}$ is the mass of a condensing molecule, $\rho_{\rm s}$ is the saturation vapor density, $S\equiv \rho_{\rm v}/\rho_{\rm s}$ is the saturation ratio, $\rho_{\rm v}$ is the vapor mass density, {  and $\sigma$ is the surface energy of condensed ice.
The second term in the bracket accounts for the increase of equilibrium vapor pressure on a curved surface, the so-called Kelvin effect.
}
We only take into account $\rm C_2H_4$ ice because the condensation rates of other ices are orders of magnitude lower than that of $\rm C_2H_4$ \citep{Strobel+90,Strobel&Summers95}.
{The surface energy of C$_2$H$_4$ is taken from \citet{Moses+92}, given by}
\begin{equation}
    \sigma=-2.37-0.1854(T-273.15)~{\rm erg~{cm}^{-2}}.
\end{equation}
The saturation vapor pressure of $\rm C_2H_4$ ice is given by \citep{Fray&Schmitt09_vapor-pressure}
\begin{eqnarray}\label{eq:P_s}
    \nonumber
    \log{(P_{\rm C_2H_4}~{[\rm bar]})}&=&15.40 -2.206 
    \times{10}^3T^{-1}-1.216\times{10}^4T^{-2}\\
    &&+2.843\times{10}^5T^{-3}-2.203\times{10}^6T^{-4},
\end{eqnarray}
where the temperature is expressed in Kelvin.
{ 
We simultaneously simulate the vertical distribution of C$_2$H$_4$ vapor with a diffusion equation given by
\begin{equation}\label{eq:trans_vapor}
    \frac{\partial \rho_{\rm v}}{\partial t}=\frac{\partial}{\partial z}\left[\rho_{\rm g}K_{\rm z}\frac{\partial}{\partial z}\left( \frac{\rho_{\rm v}}{\rho_{\rm g}} \right)\right]-\int_{\rm 0}^{\infty}\left( \frac{dm}{dt}\right)_{\rm cond}n(m)dm + \mathcal{S}(z),
\end{equation}
where $\mathcal{S}(z)$ is the net production rate of the condensing vapor, which we approximate by the Gaussian
\begin{equation}
    \mathcal{S}(z)=\frac{F_{\rm vap}}{\sigma_{\rm z}\sqrt{2\pi}}\exp{\left( \frac{(z-z_{\rm 0})^2}{2\sigma_{\rm z}^2}\right)},
\end{equation}
where $F_{\rm vap}$ is the column-integrated production rate of C$_2$H$_4$ vapor, $z_{\rm 0}$ is the altitude where the vapor is predominantly produced, and $\sigma_{\rm z}$ is the width of the distribution.
We set $z_{\rm 0}=20~{\rm km}$ and $\sigma_{\rm z}=5~{\rm km}$ to mimic the photochemical production of C$_2$H$_4$ \citep{Strobel+90}.
The $K_{\rm z}$ is the eddy diffusion coefficient and set to $K_{\rm z}=4\times{10}^{3}~{\rm {cm}^2~s^{-1}}$ following \citet{Krasnopolsky&Cruikshank95}.}
For the ice balls, the particle density is set to $\rho_{\rm p}=0.64~{\rm g~{cm}^{-3}}$, the material density of C$_2$H$_4$ ice \citep[][]{Satorre+17}.

\subsection{Numerical Procedures for the Microphysical Model}
We numerically solve the master equations from the surface to $250~{\rm km}$ until the system reaches a steady state.
The master equations are Equation \eqref{eq:basic} for the ice-free spheres, Equations \eqref{eq:basic} and \eqref{eq:basic2} for the ice-free and ice aggregates, and Equation \eqref{eq:basic_cond} for the ice balls (see Table \ref{table:1}).
The mass coordinate is divided into linearly spaced bins, $m_{\rm k}=km_{\rm 0}$, for $m_{\rm k}<N_{\rm bd}m_{\rm 0}/2$ and logarithmically spaced bins, $m_{\rm k}=m_{\rm k-1}{10}^{1/N_{\rm bd}}$, for $m_{\rm k}\geq N_{\rm bd}m_{\rm 0}/2$, where we adopt the mass resolution of $N_{\rm bd}=5$, i.e., $m_{\rm k}/m_{\rm k-1}{\approx} 1.58$.
{  Here $m_{\rm 0}=4\pi r_{\rm mon}^3\rho_{\rm p}/3$ is the smallest mass grid, corresponding to the monomer mass.}
We assume that the initial haze particles are produced at the ionosphere, as recognized for Titan hazes.
We impose the downward mass flux at the top boundary, $z=250~{\rm km}$, that is close to the bottom of the ionosphere \citep{Tyler+89}.
Haze particles are freely settling out onto the surface.
{In the ice ball model, we set zero vapor fluxes at both upper and lower boundaries.}
The vertical pressure-temperature profile is taken from a radiative-conductive model of \citet{Strobel&Zhu17}. 
The downward mass flux $F_{\rm top}$, the monomer size $r_{\rm mon}$, and the charge density $q_{\rm eq}$ are free parameters.
{The column-integrated C$_2$H$_4$ production rate $F_{\rm vap}$ is an additional parameter in the ice ball model.}
We select the mass flux comparable to the column-integrated Ly$\alpha$ photolysis rate of CH$_4$, $4$--$8\times{10}^{-15}~{\rm g~{cm}^{-2}~s^{-1}}$ \citep{Strobel+90,Strobel&Summers95,Bertrand&Forget17}.

{We note the caveat regarding the interpretation on downward mass flux $F_{\rm top}$ in the ice aggregate model.
While $F_{\rm top}$ in the ice-free and ice ball models expresses the mass flux of ice-free particles, $F_{\rm top}$ in the ice aggregate model is the sum of ice-free and condensed ice mass flux.
In other words, we do not disentangle the ice-free and condensed ice components in the $F_{\rm top}$ used for the ice aggregate model. Instead, we assume ice-coated monomers without solving the ice condensation onto aggregates. A future study with ice condensation growth of aggregates could better constrain the ice-free and condensed ice mass flux in the ice aggregate model, respectively.
}

\subsection{Calculations of Observational Signatures}\label{sec:method_observation}

\ko{We constrain each haze formation scenario based on the observations of Voyager 2.}
We first compare the model results with the extinction coefficient $\alpha_{\rm ext}$ of the Triton haze constrained by UV solar occultation observations \citep{Herbert&Sandel91,Krasbopolsky+92}. 
The extinction coefficient is calculated by
\begin{equation}
    \alpha_{\rm ext}(z)=\int_{\rm 0}^{\rm \infty}(\sigma_{\rm sca}+\sigma_{\rm abs})n(m,z)dm,
\end{equation}
where $\sigma_{\rm sca}$ and $\sigma_{\rm abs}$ are the scattering and absorption cross sections, respectively.
\ko{\citet{Krasbopolsky+92} reported that the extinction coefficient is $\alpha_{\rm ext}\sim{10}^{-9}$--${10}^{-8}~{\rm {cm}^{-1}}$ at $\lambda=0.15~{\rm {\mu}m}$ from $z=0$--$30~{\rm km}$.}

\ko{The visible photometric observations \citep[e.g.,][]{Hillier+90,Hillier+91,Rages&Pollack92} are also useful to investigate the haze formation process. 
We use scattered-light intensity from Triton hazes constrained by disk-resolved observations \citep{Rages&Pollack92} because disk-averaged observations are highly contaminated by the discrete clouds \citep{Hillier&Veverka94}. 
Assuming optically thin hazes, the scattered-light intensity, $I$, is calculated as}
\begin{equation}\label{eq:IF_slant_calc}
    \frac{I(\theta,R)}{F}\approx \int_{\rm R}^{\infty} \int_{\rm 0}^{\infty} \frac{P(\theta,m,R')\sigma_{\rm sca}n(m,R')R'}{4\sqrt{R'^2-R^2}}dmdR',
\end{equation}
where $F$ is the incident flux, $P(\theta)$ is the scattering phase function, $\theta$ is the scattering angle, and $R$ is the radial distance from the center of Triton.
When $P(\theta)$ and $\sigma_{\rm sca}$ are spatially homogeneous and $n(m,r)$ is proportional to atmospheric density, Equation \eqref{eq:IF_slant_calc} reduces to the $I/F\approx P(\theta)\tau_{\rm s,chord}/4$ used in \citet{Cheng+17}, where $\tau_{\rm s,chord}$ is the scattering chord optical depth. 
Note that the phase function is normalized to $\int P(\theta) d\Omega = 4\pi$.
The phase angle is ${180}^{\circ}-\theta$; thus, small scattering angles correspond to high phase angles.

{We utilize a haze profile retrieved by \citet{Rages&Pollack92} from the spatially resolving photometric observations at high phase angles.
Although low phase angle observations are available as well \citep{Pollack+90}, we do not use them because low phase angle observations mostly trace the reflected light from the ground rather than scattered light from hazes \citep[e.g.,][]{Hillier&Veverka94}.}
\citet{Rages&Pollack92} reported a scattered-light intensity of $I/F{\sim}0.05$--$0.1$ at $\lambda=0.431$--$0.596~{\rm {\mu}m}$ and phase angles of $140^{\circ}$--$160^{\circ}$ for a cloudless region (${15}^\circ$S, ${275}^{\circ}$E).
{  They retrieved the haze vertical profile assuming a number density varying exponentially with altitude, i.e.,
\begin{equation}\label{eq:Rages1}
    n(r,z)=n(r,0)\exp{\left( -\frac{z}{H_{\rm h}}\right)},
\end{equation}
where $n(r,z)dr$ is the number density of particles with radii between $r$ and $r+dr$ at an altitude of $z$, and $H_{\rm h}$ is the haze scale height.
They fit their models to the observations assuming a Hansen-Hovenier size distribution \citep{Hansen&Hovenier74}, given by
\begin{equation}\label{eq:Rages2}
    n(r,0)=\frac{N_{\rm h}(r_{\rm h}b_{\rm eff})^{(2b_{\rm eff}-1)/b_{\rm eff}}}{H_{\rm h}\Gamma[(1-2b_{\rm eff})/b_{\rm eff}]}r^{(1-3b_{\rm eff})/b_{\rm eff}}\exp{\left( -\frac{r}{r_{\rm h}b_{\rm eff}}\right)},
\end{equation}
where $N_{\rm h}$ is the column number density of haze particles; $r_{\rm h}$ is the cross-section-weighted mean radius; $b_{\rm eff}$ is the effective variance, which is fixed to $b_{\rm eff}=0.05$; and $\Gamma(x)$ is the gamma function.
They retrieved $N_{\rm h}$, $H_{\rm h}$, and $r_{\rm h}$ assuming the particle refractive indices of $(n,k)=(1.44,0)$. 
The retrieved parameters at the cloudless region of ($15^{\circ}$S, $275^{\circ}$E) are summarized in Table \ref{table:2}.
We calculate $I/F$ based on the haze profile described by Equations \eqref{eq:Rages1}, \eqref{eq:Rages2}, and Table \ref{table:2} using the same optical constants.
Then, we compare it with the $I/F$ calculated from our simulation results.
This approach allows us to perform model comparisons focusing only on a haze component in the scattered light.
Note that our calculated $I/F$ neglects the scattered light from the ground.
\citet{Rages&Pollack92} retrieved the haze properties assuming the photometric ground properties constrained by \citet{Hillier+91}.
Thus, our analysis implicitly assumes that scattered light from the ground is represented as in \citet{Hillier+91}.
}
The $I/F$ varies with altitude (see Equation \ref{eq:IF_slant_calc}), and we perform the model comparison at the surface level ($R=R_{\rm s}$), where the radius of Triton is $R_{\rm s}=1353.4~{\rm km}$.

\begin{table}[t]
\begin{threeparttable}
  \caption{Haze properties retrieved by \citet{Rages&Pollack92}.}\label{table:2}
  \centering
  \begin{tabular}{l r} \hline
    Parameter & Retrieved value\tnote{a} \\\hline \hline
    Column number density $N_{\rm h}$& ($2.0\pm0.6)\times{10}^{6}~{\rm {cm}^{-2}}$\\
     Haze scale height $H_{\rm h}$& $11.0\pm0.6~{\rm {km}}$\\
     Cross-section averaged radius $r_{\rm h}$& $0.173\pm0.012~{\rm \mu{m}}$\\ \hline
  \end{tabular}
      \begin{tablenotes}
        \raggedright
        \item[a] Retrieved at a cloudless region ($15^{\circ}S$, $275^{\circ}E$) with optical constants of $(n,k)=(1.44,0)$.
    \end{tablenotes}
 \end{threeparttable} 
\end{table}

\ko{We calculate the optical properties of the aggregates using the modified mean-field theory code of \citet{Tazaki&Tanaka18}.
The modified mean-field theory computes optical properties of fractal aggregates by means of a mean-field approach \citep{Berry&Percival86}. 
Recently, \citet{Tazaki&Tanaka18} modified the mean-field theory formulated by \citet{Botet+97} and \citet{Rannou+97} to improve the erroneous behavior of the single-scattering albedo when multiple scattering becomes important.}
We adopt the Gaussian cutoff for the two-point correlation function specifying monomer configurations \citep{Tazaki+16}.
For the spheres, we apply the Mie theory code of \citet{Bohren&Huffman83}.
For ice-free hazes, we adopt the complex refractive index of the Titan tholin \citep{Khare+84}, which was also used for Pluto's hazes \citep{Gladstone+16,Gao+17}. 
For the ice balls and aggregates, we use the refractive index of C$_2$H$_4$ ice.
There is no published refractive index of C$_2$H$_4$ ice that is available for a UV wavelength of $\lambda=0.15~{\rm \mu m}$.
Hence, we assume the real refractive index of $1.48$ measured at visible \citep[$0.633~{\rm \mu m}$,][]{Satorre+17} and the imaginary refractive index of zero, constant for all wavelengths.
{  We will discuss the sensitivity of our results to the assumed optical constants in Section \ref{sec:test_k}.}
Future laboratory studies of optical constants of hydrocarbon ices are greatly needed to help assess the optical properties of icy hazes.

\begin{figure*}[t]
\centering
\includegraphics[clip, width=\hsize]{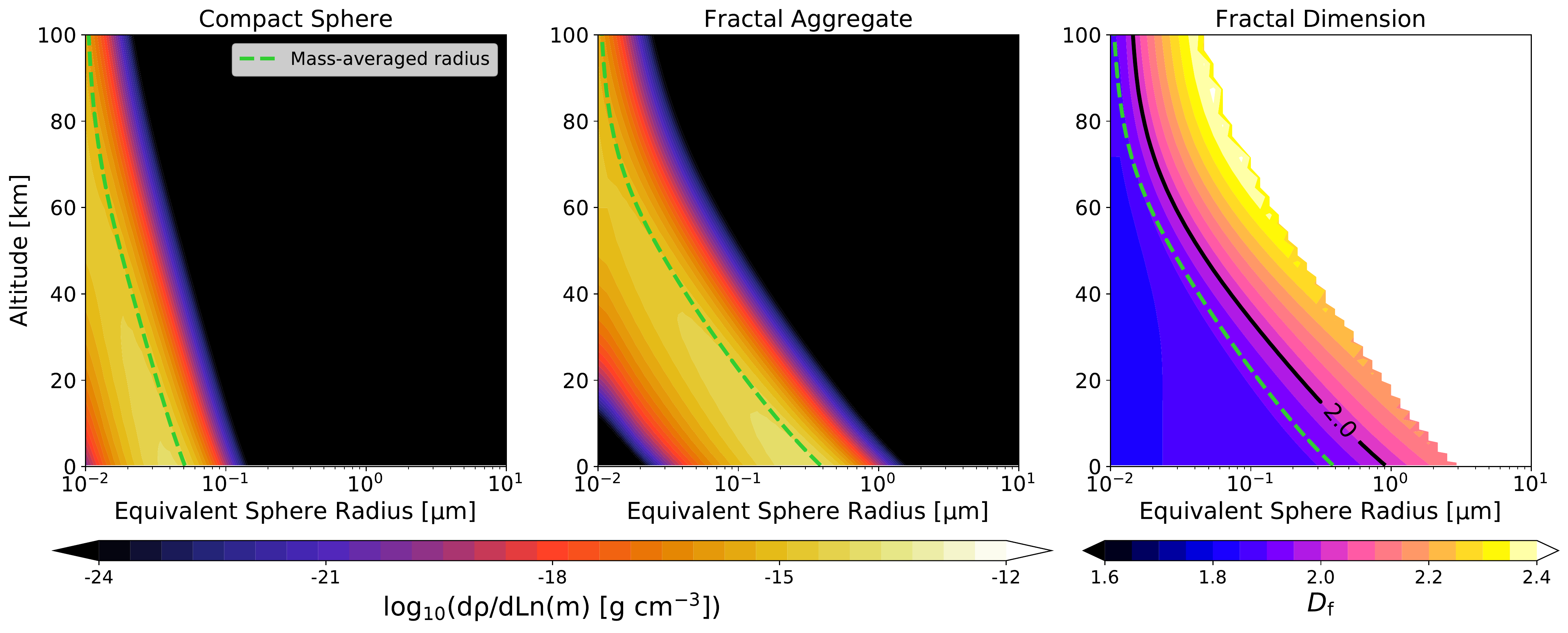}
\caption{Vertical size and porosity distributions of hazes on Triton. The vertical and horizontal axes are altitude and the mass-equivalent sphere radius, i.e., $r_{\rm mon}N_{\rm mon}^{1/3}$, respectively. The left and middle panels show the size distributions in $d(mn)/d\log{(m)}$ (colorscale) for sphere and aggregate cases, respectively. The right panel shows the fractal dimension (colorscale) in each mass and vertical grid derived by Equation \eqref{eq:Df}, where the green doted line denotes the equivalent sphere radius corresponding to the mass-averaged mass (Equation \ref{eq:mp}).
We set the downward mass flux, monomer size, and particle charge to $F_{\rm top}=3\times{10}^{-15}~{\rm g~{cm}^{-2}~s^{-1}}$, $r_{\rm mon}=10~{\rm nm}$, and $q_{\rm e}=0$, respectively.}
\label{fig:Triton_result0}
\end{figure*}
\begin{figure*}[t]
\centering
\includegraphics[clip, width=0.95\hsize]{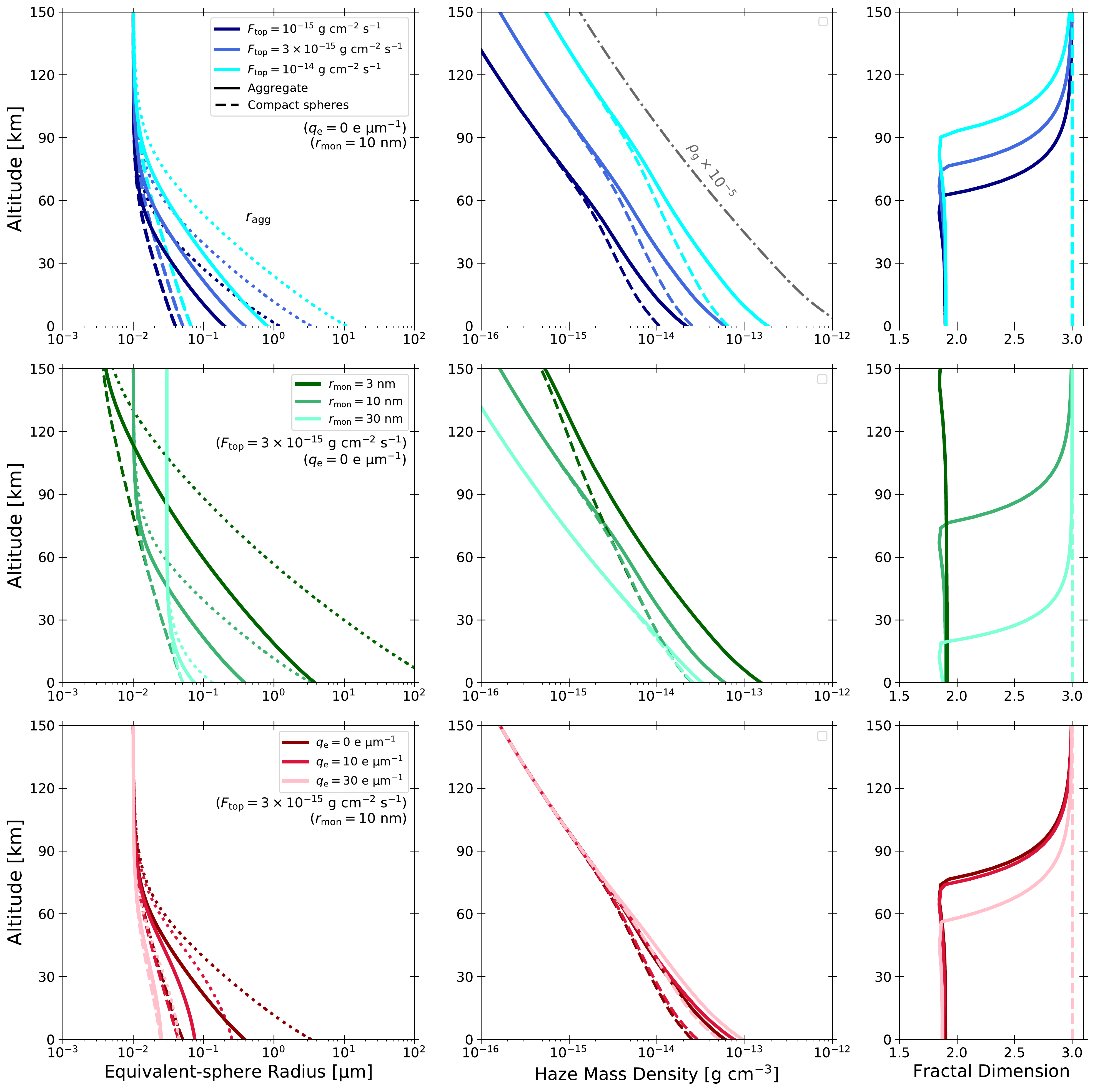}
\caption{Sensitivity study of vertical haze profiles on essential model parameters. From top to bottom, each row exhibits the profiles for the different downward mass flux, monomer size, and charge density, respectively.
The different colored lines show the profiles for different $F_{\rm top}$ in the top row, $r_{\rm mon}$ in the middle row, and $q_{\rm e}$ in the bottom row, respectively.
From left to right, each column shows the mass-averaged equivalent sphere radius, haze mass density, and mass-averaged fractal dimension, respectively. 
Solid and dashed lines exhibit the profiles for the aggregates and spheres, respectively. 
The dotted lines in the left column denote the mass-averaged characteristic radius of aggregates.}
\label{fig:vertical_profiles}
\end{figure*}

\section{Haze Formation without Ice Condensation}\label{sec:result1}

\subsection{Haze Vertical Profiles}\label{sec:vertical_icefree}
We begin by investigating the haze formation without ice condensation. 
Figure \ref{fig:Triton_result0} shows the vertical size distributions of the haze particles.
\ko{
The particles are larger at lower altitudes because of a slower settling velocity at a lower atmosphere with a higher atmospheric density.
The longer settling timescale enables particles to grow into larger sizes before they settle down.
The spheres start to grow gradually below $z{\sim}100~{\rm km}$.
The particle sizes reach ${\sim}0.05~{\rm \mu m}$ below $30~{\rm km}$, where hazes were observed by Voyager 2.
}

\ko{In general, the aggregates can grow into sizes much larger than those of the spheres.
The aggregates also start to grow below $z{\sim}100~{\rm km}$ and eventually become particles with mass-equivalent sphere radii of ${\sim}0.5~{\rm \mu m}$ near the ground. 
The mass-equivalent sphere radius is a metric of a particle mass and defined as
\begin{equation}
    r_{\rm eq}\equiv r_{\rm mon}N_{\rm mon}^{1/3}.
\end{equation}
Because of their lower bulk density, aggregates have slower settling velocities and grow more efficiently than the spheres. 
}

\ko{Our simulations yield a fractal dimension of the aggregates close to $D_{\rm f}=2$.}
The right panel of Figure \ref{fig:Triton_result0} shows the fractal dimension as a function of the equivalent sphere radius and altitude.
Larger aggregates have higher fractal dimensions because they are prone to experience collisions with large size ratios.
The fractal dimension ranges from $D_{\rm f}\approx 1.8$ to $2.2$, indicating that the growth is mainly driven by binary collisions between similar-sized particles. 
To see a typical value of $D_{\rm f}$, we introduce the mass-averaged mass:
\begin{equation}\label{eq:mp}
    \overline{m}_{\rm p}\equiv \frac{\int m^2ndm}{\int mndm}.
\end{equation}
\ko{
The averaged mass approximately traces the peak of the mass distribution \citep[][see also green dashed lines in Figure \ref{fig:Triton_result0}]{Ormel&Spaans08}.
As shown in the right panel of Figure \ref{fig:Triton_result0}, the mass-dominating aggregates have the fractal dimension of $D_{\rm f}\approx 1.9$.
Our results are in good agreement with the fractal dimension $D_{\rm f}\approx1.8\pm0.2$ found by full $N$-body simulations of Brownian-motion-driven coagulation \citep{Kempf+99}.
}

\begin{figure*}[t]
\centering
\includegraphics[clip, width=\hsize]{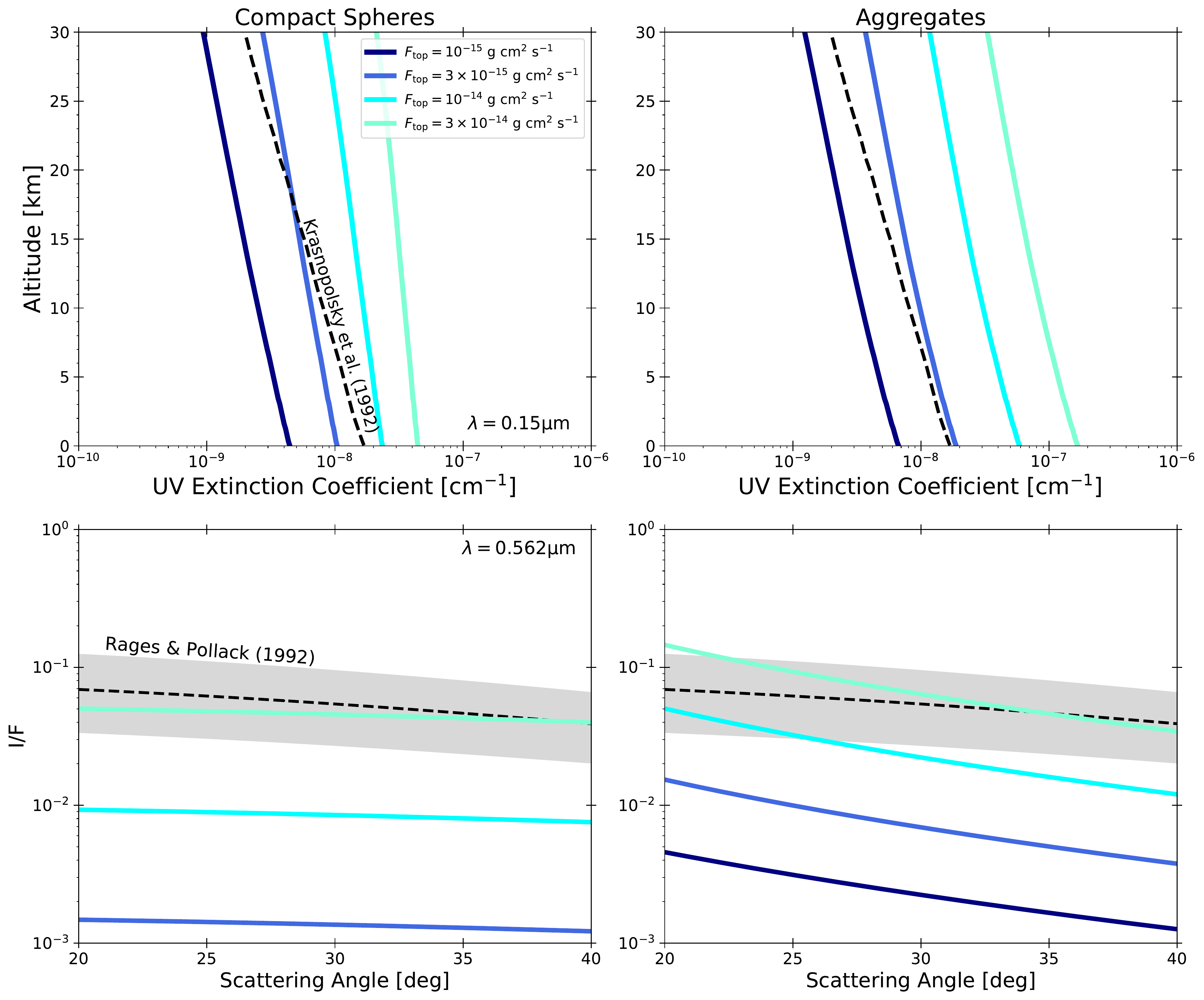}
\caption{UV ($\lambda=0.15~{\rm \mu m}$) extinction coefficient at each altitude (top) and visible ($\lambda=0.562~{\rm {\mu}m}$) scattered light intensity at the ground (bottom) of the simulated ice-free hazes. 
The intensity is expressed by the ratio to the incident flux, $I/F$, as a function of scattering angle. 
The different colored lines indicate the results for different haze mass flux, where we fix the monomer sizes and particle charge to $r_{\rm mon}=10~{\rm nm}$ and $q_{\rm e}=0$. 
The left and right columns show the results for the spheres and aggregates, respectively. 
The black dashed lines in the top panels denote the extinction coefficient retrieved by \citet{Krasbopolsky+92}, while the dashed lines in the bottom panels show the $I/F$ calculated by the haze profiles retrieved by \citet{Rages&Pollack92} with $1\sigma$ uncertainties (gray shaded regions). }
\label{fig:comparison_Flux}
\end{figure*}

\ko{The haze vertical distributions substantially change with the mass flux, monomer size, and particle charge.}
Figure \ref{fig:vertical_profiles} shows the vertical distributions of the mass-averaged size, haze mass density, and the fractal dimension of mass-dominant particles for different mass flux (top row), monomer size (middle row), and particle charge (bottom row).
The higher mass flux leads to larger particle sizes and higher mass density.
The aggregates grow into the equivalent sphere radius of ${\sim}0.3$--$0.8~{\rm \mu m}$ near the ground, while the spherical particles only grow into the smaller size of ${\sim}0.03$--$0.06~{\rm \mu m}$.
We note that the characteristic radii of the fractal aggregates $r_{\rm agg}$ are much larger than the equivalent sphere radii (dotted lines).
For a given mass flux, the aggregates always have larger sizes and higher mass density than those of the spheres due to slower settling velocities.

\ko{The fractal dimension of the aggregates always approaches $D_{\rm f}\approx1.9$ at low altitudes.}
In the upper atmosphere, the aggregates have a fractal dimension of $D_{\rm f}\approx3$ because the mass-dominant monomers settle without collisional growth.
As they grow, the fractal dimension of aggregates approaches $\approx 1.9$, independent of mass flux, monomer size, and particle charges.
Our calculated fractal dimension for Triton hazes is similar to that recognized for Titan hazes \citep{Rannou+97}. Because the fractal dimension is almost 2, \ko{the mass density of the aggregates is nearly proportional to atmospheric density.} 
In the steady state, the mass density is given by the mass conservation,
\begin{equation}\label{eq:rho_haze}
    \rho_{\rm haze}=\frac{F_{\rm top}}{\overline{v_{\rm t}}}
\end{equation}
where $\overline{v_{\rm t}}$ is the mass-averaged settling velocity, defined as
\begin{equation}\label{eq:vtbar}
    \overline{v_{\rm t}}\equiv\frac{\int v_{\rm t}mn(m)dm}{\int mn(m)dm}\sim \frac{g\rho_{\rm 0}r_{\rm mon}^{3-D_{\rm f}}}{\rho_{\rm g}C_{\rm s}}\frac{\int r_{\rm agg}^{D_{\rm f}-2}mn(m)dm}{\int mn(m)dm},
\end{equation}
where $\rho_{\rm 0}$ is the material density, and we have used Equations \eqref{eq:vt} with $A\sim\pi r_{\rm agg}^2$ and the relation of $\rho_{\rm p}=\rho_{\rm 0}(r_{\rm agg}/r_{\rm mon})^{D_{\rm f}-3}$.
One can see that the settling velocity is invariant with aggregate sizes for $D_{\rm f}=2$, as the mass-to-area ratio is constant.
As a result, the settling velocity is inversely proportional to $\rho_{\rm g}$, and thus the mass density is proportional to $\rho_{\rm g}$  {  \citep[see e.g.,][]{Cabane+92,Cabane+93}}.
On the other hand, the mass density of the spheres is not proportional to $\rho_{\rm g}$ because the particle mass-to-area ratio varies with altitude as the particle grows.

\ko{The monomer size mainly affects the vertical distributions of the aggregates (middle row of Figure \ref{fig:vertical_profiles}).}
For the spheres, the vertical distributions are insensitive to the monomer size, as the particle size is controlled by a balance between collision and settling timescales regardless of initial sizes. 
For the aggregates, larger monomers lead to smaller particle sizes and lower mass densities. 
This is because the settling velocity for a larger monomer is higher and inhibits particle growth.

The particle charge affects the aggregate sizes but does not affect the mass density (bottom row of Figure \ref{fig:vertical_profiles}).
The aggregates grow into the equivalent sphere radii of ${\sim}0.5~{\rm \mu m}$ for the no-charge case, while they grow into the radii of only ${\sim}0.1~{\rm \mu m}$ for $q_{\rm e}=10~{\rm e~{{\mu}m}^{-1}}$ and ${\sim}0.03~{\rm \mu m}$ for $q_{\rm e}=30~{\rm e~{{\mu}m}^{-1}}$.
Nevertheless, the mass density is nearly invariant with the charge density.
This is because the reducing aggregate sizes does not affect the settling velocity for $D_{\rm f}\approx2$ (see Equation \ref{eq:vtbar}).
The charge effect is almost negligible for the spheres because they hardly grow sufficiently large for the electrostatic repulsion to be important.
In the next section, we will first use the model results to explain the UV extinction profile and then match the visible scattering observations.
As we will show, ice-free hazes could not simultaneously match both observations if we assume that they are absorbing materials such as Titan's tholin.

\begin{figure}[t]
\centering
\includegraphics[clip, width=\hsize]{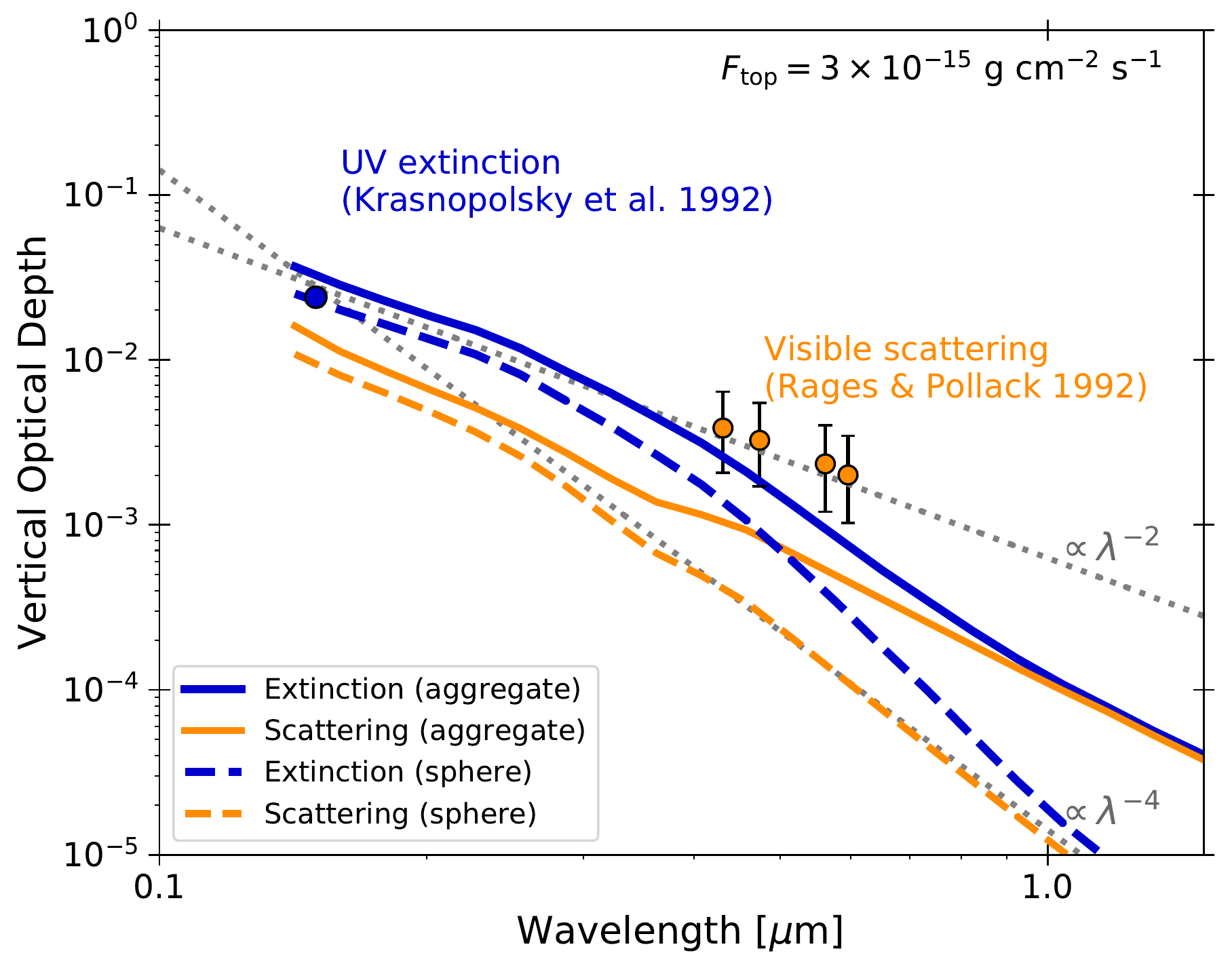}
\caption{Vertical optical depth of the simulated haze as a function of wavelength. 
The blue and orange lines show the extinction and scattering optical depth, respectively.
The solid and dashed lines denote the optical depth for the aggregates and spheres, respectively.
The haze mass flux, monomer sizes, and particle charge are set to $F_{\rm top}=3\times{10}^{-15}~{\rm g~{cm}^{-2}~s^{-1}}$, $r_{\rm mon}=10~{\rm nm}$, and $q_{\rm e}=0$, respectively.
The blue dot indicates the extinction optical depth reported by \citet{Krasbopolsky+92}, while the orange dots indicate the scattering optical depth reported by \citet{Rages&Pollack92}.
The error bars in the data of \citet{Rages&Pollack92} are estimated from $1\sigma$ uncertainties in their derived haze properties (see also Table \ref{table:2}).
The gray dotted lines denote the proportionality of $\lambda^{-2}$ and $\lambda^{-4}$.
}
\label{fig:tau_schematic}
\end{figure}

\subsection{Comparisons with Voyager 2 Observations}\label{sec:UV}
\ko{The aggregates generally explain the UV observations better than the spheres do.}
The top panels of Figure \ref{fig:comparison_Flux} show the vertical extinction profiles for different haze mass flux. 
We set the monomer sizes and particle charge to $r_{\rm mon}=10~{\rm nm}$ and $q_{\rm e}=0$.
The calculated extinction profile is approximately proportional to the mass flux. 
The aggregates explain the retrieved extinction coefficient well in the case of $F_{\rm top}{\sim}3\times{10}^{-15}~{\rm g~{cm}^{-2}~s^{-1}}$.
Although the spheres can crudely explain the \ko{magnitude} of the extinction coefficient for $F_{\rm top}=3$--$10\times{10}^{-15}~{\rm g~{cm}^{-2}~s^{-1}}$, the model fails to explain the vertical gradient.
This is because the spheres settle faster than the aggregates and decrease the mass density in the lower atmosphere.
This leads to an extinction coefficient for the spheres that is smaller than the coefficient for the aggregates in the lower atmosphere.

\ko{Comparing the model with the visible observations, however, we find that there is no solution that explains both visible and UV observations simultaneously.}
The bottom panels of Figure \ref{fig:comparison_Flux} show the visible $I/F$ and that calculated by the haze profiles of \citet[][]{Rages&Pollack92}.
In general, the $I/F$ increases with increasing the mass flux because a higher haze mass density yields a higher scattering optical depth.
The aggregates explain the $I/F$ in the observed scattering angles from ${20}^{\circ}$ to ${40}^{\circ}$ with a mass flux of $F_{\rm top}{\sim}1$--$3\times{10}^{-14}~{\rm g~{cm}^{-2}~s^{-1}}$.
On the other hand, the spheres match the $I/F$ with a higher mass flux of $F_{\rm top}\sim3\times{10}^{-14}~{\rm g~{cm}^{-2}~s^{-1}}$. 
Both spheres and aggregates require a mass flux an order of magnitude higher than that needed to explain the UV extinction coefficient.

\ko{
The discrepancy between the required mass fluxes to match the UV and visible observations originates from the wavelength dependence of haze opacity.}
To illustrate this, Figure \ref{fig:tau_schematic} shows the extinction and scattering optical depth of spherical and aggregate hazes as a function of wavelength. 
Also shown are the observed extinction optical depth in UV \citep{Krasbopolsky+92} and the scattering optical depth in visible \citep{Rages&Pollack92}.
The observed optical depths are nearly proportional to $\lambda^{-2}$ from the UV extinction to visible scattering. Note that the extinction optical depth cannot be smaller than the scattering optical depth at the same wavelength. Therefore, the observations imply that the extinction optical depth should probably change with the wavelength in a shallower slope than $\lambda^{-2}$.

Neither spheres nor aggregates, if made of absorbing materials such as Titan's tholin, can produce this wavelength dependence (Figure \ref{fig:tau_schematic}). In our ice-free sphere case, because small particle sizes induce Rayleigh scattering, the spheres show the scattering optical depth nearly proportional to $\lambda^{-4}$ and thus cannot explain the observations.
\ko{In the aggregate scenario, the scattering cross section of the aggregate with $D_{\rm f}\approx2$ actually follows $\approx \lambda^{-2}$ (see Appendix \ref{appendix:anal})}.
However, because the materials are absorbing (i.e., low single-scattering albedo), the UV extinction optical depth is much larger than the scattering optical depth. 
As a result, the wavelength dependence from the UV extinction to visible scattering is steeper than $\lambda^{-2}$.
We note that different monomer sizes and particle charge cannot reconcile the discrepancy, as demonstrated in Appendices \ref{appendix:anal}, \ref{appendix:monomer}, and \ref{appendix:charge}.
In other words, unless the aggregates were made of conservatively scattering materials (such as very bright ices) rather than absorbing materials (like Titan's tholin), they should not be able to explain both the UV and visible data. This argument motivates us to invoke the ice condensation scenario below.

\section{Ice Condensation as a Nature of Triton Haze}\label{sec:discussion}

We suggest that ice condensation on haze particles plays a critical role in haze formation on Triton.
The condensation induces further growth of haze particles that can suppress Rayleigh scattering.
The condensation also alters the refractive index of haze particles that may increase the single-scattering albedo.
Triton's atmosphere is so cold that a number of hydrocarbon ices can be condensed into solid \citep{Strobel+90,Strobel&Summers95}.
Thus, ice condensation can naturally resolve the model--data discrepancy in ice-free absorbing hazes.
In what follows, we demonstrate that both ice balls and ice aggregates can successfully explain the observations of Triton hazes and how future observations could distinguish the two scenarios.

\begin{figure}[t]
\centering
\includegraphics[clip, width=\hsize]{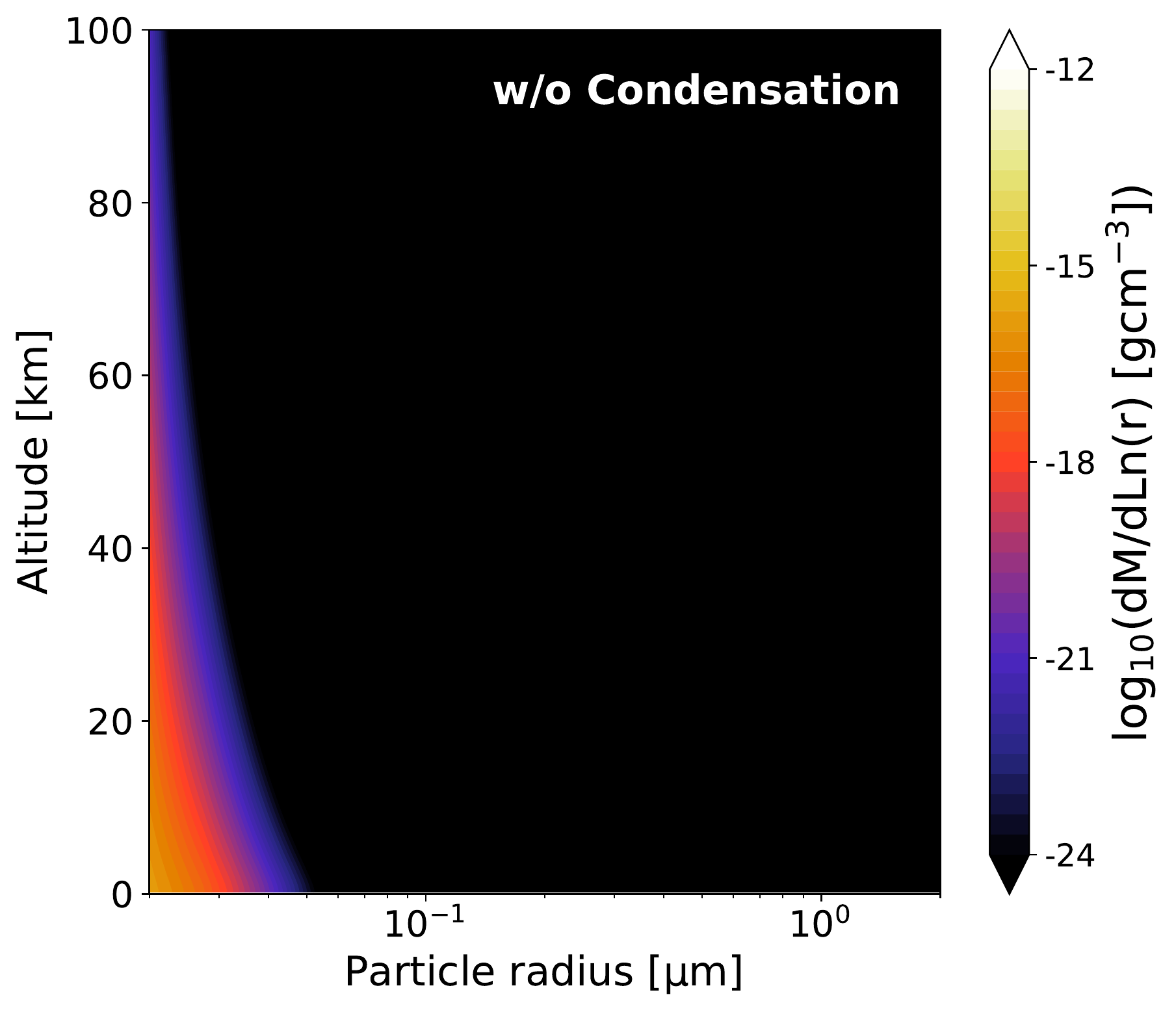}
\includegraphics[clip, width=\hsize]{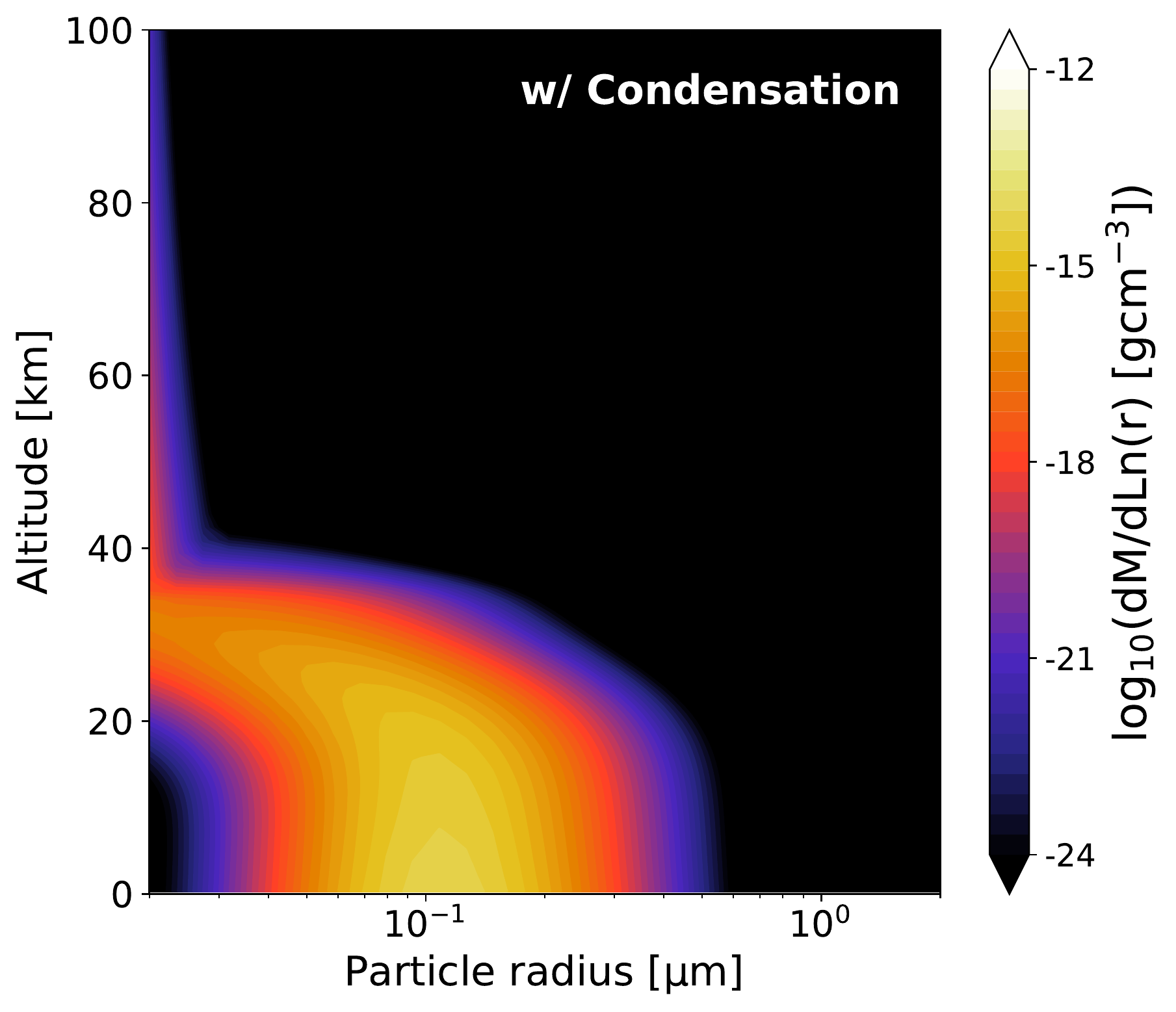}
\caption{Vertical size distributions of the ice balls. The top and bottom panel shows the distributions without and with condensation of $\rm C_2H_4$ ices, respectively. The downward ice-free mass flux is set to $F_{\rm top}= 10^{-17}~{\rm g~{cm}^{-2}~s^{-1}}$.
{The column-integrated C$_2$H$_4$ production rate is set to $F_{\rm vap}= 3\times10^{-15}~{\rm g~{cm}^{-2}~s^{-1}}$.}
} 
\label{fig:with_condensation}
\end{figure}
\begin{figure*}[t]
\centering
\includegraphics[clip, width=\hsize]{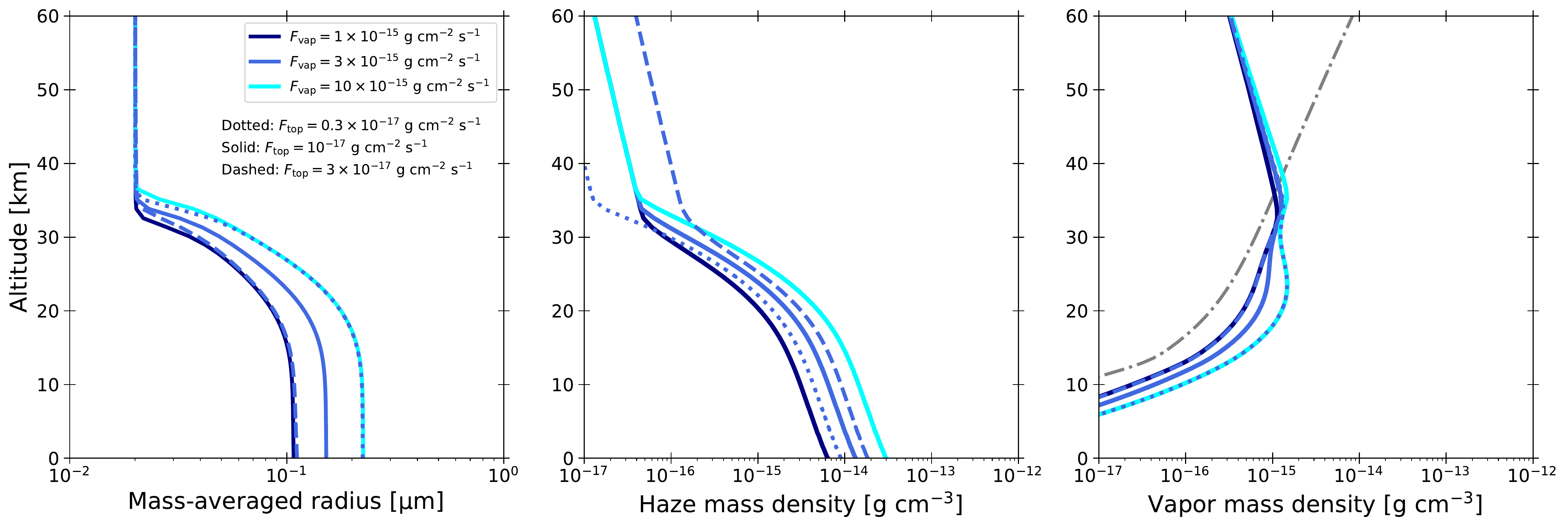}
\caption{Sensitivity study of vertical haze profiles on model parameters for the ice ball model. From left to right, each panel shows the mass-averaged particle radius, particle mass density, and vapor mass density, respectively.
The navy, blue, and cyan lines show the profiles for the column-integrated C$_2$H$_4$ production rate of $F_{\rm vap}={10}^{-15}$, $3\times{10}^{-15}$, and ${10}^{-14}~{\rm g~{cm}^{-2}~s^{-1}}$, respectively. { The downward ice-free mass flux is set to $F_{\rm top}={10}^{-17}~{\rm g~{cm}^{-2}~s^{-1}}$.}
The blue dotted and dashed lines show the profiles for $F_{\rm vap}=3\times{10}^{-15}~{\rm g~{cm}^{-2}~s^{-1}}$ with different ice-free downward flux of $F_{\rm top}=0.3$ and $3\times{10}^{-17}~{\rm g~{cm}^{-2}~s^{-1}}$. In the right panel, the gray dash-dot line denotes the saturation mass density of C$_2$H$_4$.
} 
\label{fig:ice_ball_revise}
\end{figure*}
\begin{figure*}[t]
\centering
\includegraphics[clip, width=\hsize]{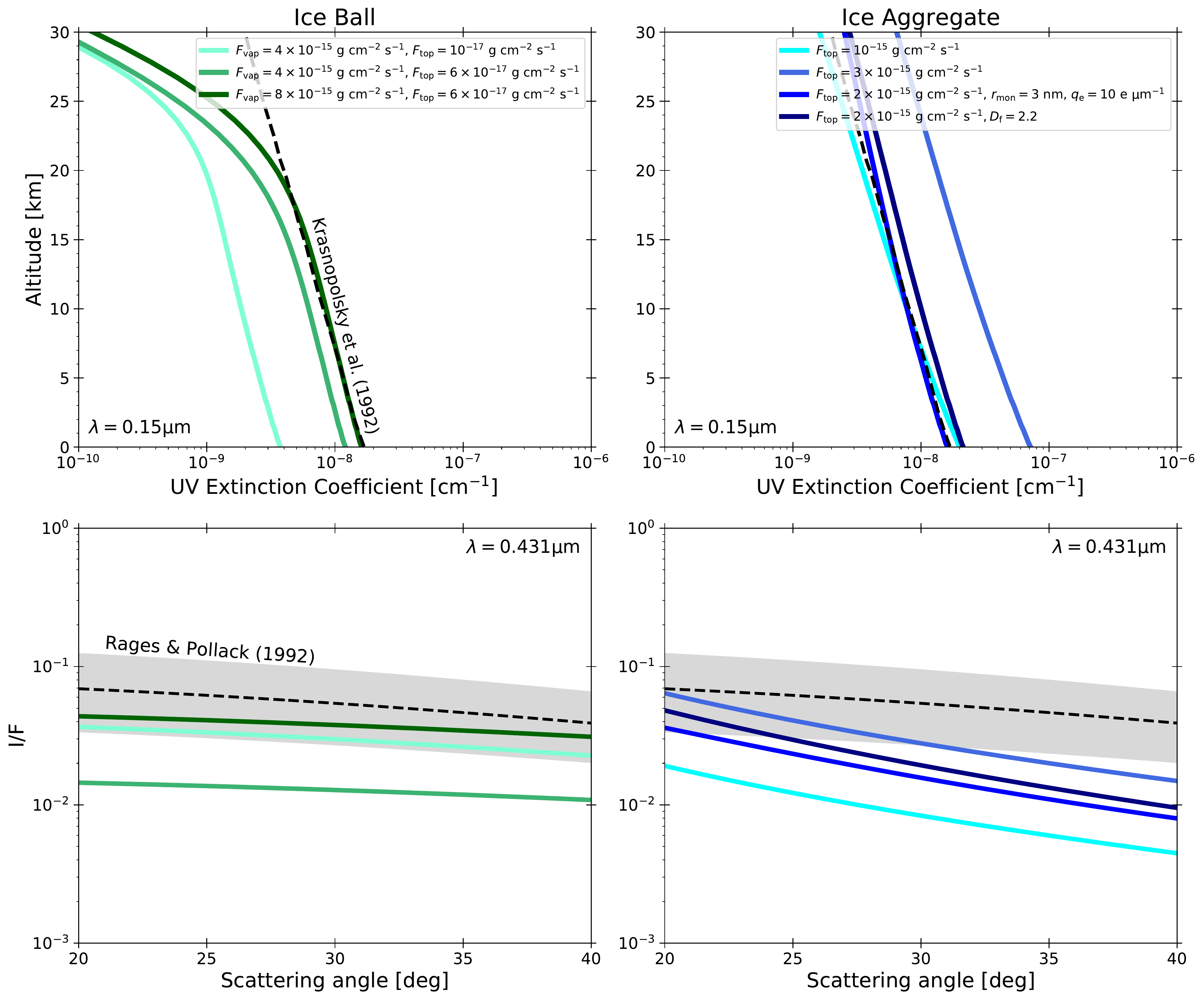}
\caption{The UV extinction coefficient (top row) and visible $I/F$ (bottom row) of icy haze models. The left and right column show the results for the ice ball and ice aggregate models, respectively. 
{For the ice balls, the light-green and green lines show the results for the column-integrated C$_2$H$_4$ production rate of $F_{\rm vap}=4\times{10}^{-15}~{\rm g~{cm}^{-2}~s^{-1}}$ with ice-free mass flux of  $F_{\rm top}={10}^{-17}$ and $6\times{10}^{-17}~{\rm g~{cm}^{-2}~s^{-1}}$, respectively. The dark-green line shows the best model of the ice balls, in which $F_{\rm vap}=8\times{10}^{-15}~{\rm g~{cm}^{-2}~s^{-1}}$ and $F_{\rm top}=6\times{10}^{-17}~{\rm g~{cm}^{-2}~s^{-1}}$.}
{  For the ice aggregates, the cyan and lighter blue lines show the results for the mass flux of $F_{\rm top}={10}^{-15}$ and $3\times{10}^{-15}~{\rm g~{cm}^{-2}~s^{-1}}$, respectively.
{ The darker blue line shows the ice aggregate model with a smaller monomer size of $r_{\rm mon}=3\ {\rm nm}$, in which $F_{\rm top}=2\times{10}^{-15}~{\rm g~{cm}^{-2}~s^{-1}}$, and $q_{\rm e}=10~{\rm e~{{\mu}m}^{-1}}$.}
The navy line shows the best model of the ice aggregates, in which the mass flux is $F_{\rm top}=2\times{10}^{-15}~{\rm g~{cm}^{-2}~s^{-1}}$ and fractal dimension is fixed to $D_{\rm f}=2.2$.
We set the monomer sizes and particle charge to $r_{\rm mon}=20~{\rm nm}$ and $q_{\rm e}=0$ for the ice balls and $r_{\rm mon}=10~{\rm nm}$ and $q_{\rm e}=0$ for the ice aggregates, { except for the darker blue line}.}
}
\label{fig:iceball}
\end{figure*}

\subsection{Ice Ball Scenario}\label{sec:iceball}
\ko{The ice condensation allows the ice-free spheres to grow much larger.}
Figure \ref{fig:with_condensation} shows the vertical size distributions of the ice balls.
{  We set the ice-free monomer size to $20~{\rm nm}$. 
The downward ice-free mass flux is set to $F_{\rm top}={10}^{-17}~{\rm g~{cm}^{-2}~s^{-1}}$, and the column-integrated vapor production rate is set to $F_{\rm vap}=3\times{10}^{-15}~{\rm g~{cm}^{-2}~s^{-1}}$.}
{We note that a higher downward mass flux leads to a higher condensation rate because of the increase of number density and the total surface area of settling particles.}
If there is no ice condensation, the particles gradually grow via collisional growth only (top panel of Figure \ref{fig:with_condensation}).
Once ice condensation occurs, the particles rapidly grow at $z\la30~{\rm km}$ where C$_2$H$_4$ is supersaturated (bottom panel of Figure \ref{fig:with_condensation}).
The particle size is nearly constant below $z\sim20~{\rm km}$ because the condensation growth is inefficient owing to the depletion of C$_2$H$_4$ near the ground (see Figure \ref{fig:PT_Triton}).
Because of the condensation, the ice balls eventually grow as large as $\sim0.1~{\rm \mu m}$ near the ground, appreciably larger than the sizes of the ice-free spheres.
The low material density of C$_2$H$_4$ ice \citep[$0.64~{\rm g~{cm}^{-3}}$,][]{Satorre+17} also facilitates the efficient particle growth by slowing down the particle settling.

{ 
The ice ball properties are mainly controlled by the vapor production rate.
Figure \ref{fig:ice_ball_revise} shows the mass-averaged radius and mass density of the ice ball particles and the C$_2$H$_4$ vapor mass density for different vapor production rates.
The higher the vapor production rate, the larger the particle sizes and the higher the mass density.
This trend is intuitively understandable, as efficient vapor production facilitates condensation growth and increases the total condensed mass.
The simulated C$_2$H$_4$ vapor profiles tend to be supersaturated below $z\sim35~{\rm km}$. 
The saturation ratio can be as high as $S{\sim}3$--$10$, depending on the vapor production rate.
The particle mass flux near the ground is identical to $F_{\rm vap}$.
This means that the C$_2$H$_4$ vapor production is balanced by the condensation onto haze particles followed by gravitational settling.
The mass flux after the ice condensation is much higher than the ice-free mass flux $F_{\rm top}$ in our simulations. 
Thus, haze particles are almost purely made of the condensed ices, as assumed in our model.

The ice ball properties also depend on the ice-free monomer mass flux $F_{\rm top}$.
A higher $F_{\rm top}$ leads to a higher haze mass density.
On the other hand, a higher $F_{\rm top}$ leads to a smaller particle size, as seen in Figure \ref{fig:ice_ball_revise}.
The decrease of the particle size stems from the fact that the high ice-free flux leads to a high number density of haze particles available for condensation.
When the number density is high, the C$_2$H$_4$ vapor is efficiently removed by condensation, reducing the saturation ratio.
Since the condensation growth rate is nearly proportional to the saturation ratio for $S\gg1$ (see Equation \ref{eq:dmdt_cond}), a high ice-free monomer flux lowers the saturation ratio and slows down the condensation growth.
}

{ 
The ice ball scenario could successfully explain the observations if {the vapor production rate is sufficiently high.}
The left column of Figure \ref{fig:iceball} shows the UV extinction coefficient and visible $I/F$ of the ice balls.
Assuming a C$_2$H$_4$ production rate of $F_{\rm vap}=4\times{10}^{-15}~{\rm g~{cm}^{-2}~s^{-1}}$ and ice-free mass flux of $F_{\rm top}={10}^{-17}~{\rm g~{cm}^{-2}~s^{-1}}$, the model explains the visible $I/F$ but underestimates the UV extinction coefficient by a factor of $\sim3$.
The underestimation of the UV extinction coefficient is caused by the low mass density of ice balls.
The ice ball model could reasonably explain the UV extinction coefficient for $F_{\rm top}=6\times{10}^{-17}~{\rm g~{cm}^{-2}~s^{-1}}$, but then the visible $I/F$ is underestimated by a factor of $\sim2$--$3$ owing to reduced particle sizes.
We find that the ice ball model explains both the UV extinction coefficient and visible $I/F$ given a C$_2$H$_4$ production rate of $F_{\rm vap}=8\times{10}^{-15}~{\rm g~{cm}^{-2}~s^{-1}}$ and $F_{\rm top}=6\times{10}^{-17}~{\rm g~{cm}^{-2}~s^{-1}}$.

We note that the C$_2$H$_4$ production rate of $F_{\rm vap}=8\times{10}^{-15}~{\rm g~{cm}^{-2}~s^{-1}}$ required for the ice ball model is comparable to the CH$_4$ photolysis rate on Triton.
The column-integrated CH$_4$ photolysis rate is $8\times{10}^{-15}~{\rm g~{cm}^{-2}~s^{-1}}$ in the summer hemisphere, whereas the rate is $4\times{10}^{-15}~{\rm g~{cm}^{-2}~s^{-1}}$ in the winter hemisphere \citep{Strobel&Summers95}.
Thus, the required vapor production rate is comparable to the CH$_4$ photolysis rate in the summer hemisphere, which seems to be consistent with the region where \citet{Rages&Pollack92} retrieved the haze profile ($15^{\circ}$S, $275^{\circ}$E)\footnote{We note that the southern hemisphere was in Triton's summer season during the Voyager flyby at 1989.}, utilized in this study.
On the other hand, the column-integrated production rate of C$_2$H$_4$ itself is $\sim4\times{10}^{-15}~{\rm g~{cm}^{-2}~s^{-1}}$, a factor of $2$ lower than the required value \citep[Table 1 of][]{Krasnopolsky&Cruikshank95}.
This may imply that, if the ice ball scenario is true, the ice balls are composed of not only C$_2$H$_4$ ice but also other hydrocarbon ices, such as C$_2$H$_2$ and C$_2$H$_6$.
}

The ice ball results appear to deviate from the observed UV extinction near $z=25$--$30~{\rm km}$ where condensation has not fully occurred yet.
{  This may imply that the production region of the C$_2$H$_4$ (or other hydrocarbons) on Triton is located higher than the $20~{\rm km}$ that was suggested by \citet{Strobel+90}.
Alternatively, the deviation near $z=25$--$30~{\rm km}$ may be caused by an extrapolated vapor pressure.}
The condensation region depends on the saturation vapor pressure of C$_2$H$_4$. In this study, we use the vapor pressure from \citet{Fray&Schmitt09_vapor-pressure}, which is verified only for $T>77.3~{\rm K}$. The validity of extrapolation to colder temperatures is still under debate.
From the C$_2$H$_4$ concentration on Pluto, as measured by the New Horizons, \citet{Wong+17} suggested that the extrapolated vapor pressure is orders of magnitude higher than the actual vapor pressure \citep[for different suggestions, see][]{Luspay-Kuti+17,Krasnopolski20}.
If this was true, the condensation region of C$_2$H$_4$ on Triton would extend to above 30 km, and the UV extinction data between 25 and 30 km from Voyager 2 could be explained as well.

\subsection{Ice Aggregate Scenario}\label{sec:iceaggregate}
\ko{The ice aggregates could also explain the observations.}
{ The right column of Figure \ref{fig:iceball} shows the UV extinction coefficient and visible $I/F$ for the aggregates composed of ice monomers. 
We assume a monomer radius of $r_{\rm mon}=10~{\rm nm}$ unless otherwise indicated}. 
The ice aggregates explain the UV extinction coefficient for {  $F_{\rm top}={10}^{-15}~{\rm g~{cm}^{2}~s^{-1}}$. }
Then, the model produces a visible $I/F$ lower than that in \citet{Rages&Pollack92} by a factor of only $2$--$3$.
The result is much better than that of ice-free hazes, which yield an order-of-magnitude discrepancy.
{ A smaller monomer size of $r_{\rm mon}=3~{\rm nm}$ yields a better model fit because most particles can grow into sizes larger than the wavelength. 
We have set $q_{\rm e}=10\ {\rm {{\mu}m}^{-1}}$ so that the particles do not grow into extremely large sizes of $\gg1~{\rm {\mu}m}$. 
However, one caveat is that icy monomers with $r_{\rm mon}<10~{\rm nm}$ could be formed under limited conditions, namely small contact angles and high saturation ratios (see Section \ref{sec:nucleation}).}

The data could be explained better if the fractal dimension of ice aggregates is larger. In our current model, collisional growth alone leads to a fractal dimension of $D_{\rm f}\sim1.9$, which causes a wavelength dependence of the opacity ${\propto}\lambda^{-2.1}$ {(see Eqs \eqref{eq:Berry} in Appendix \ref{appendix:anal})} that is slightly stronger than { the dependence ${\propto}\lambda^{-2}$ of the observations.} 
{On the other hand,} aggregates with $D_{\rm f}>2$ show a ${\propto}\lambda^{-2}$ dependence { that better explains the observations}.
In reality, ice condensation may cause an increase of the fractal dimension by filling some of the pores within an aggregate.
Assuming a fractal dimension of $D_{\rm f}=2.2$, we find that the ice aggregates explain both the UV extinction coefficient and visible $I/F$ reasonably well.
The mass flux assumed in the best model is $F_{\rm top}=2\times{10}^{-15}~{\rm g~{cm}^{-2}~s^{-1}}$, 
comparable to the production rate of condensable hydrocarbons suggested by photochemical calculations, $4$--$8\times{10}^{-15}~{\rm g~{cm}^{-2}~s^{-1}}$ \citep[][]{Strobel+90,Strobel&Summers95}.

\subsection{Implications for future observations}\label{sec:future}
\begin{figure}[t]
\centering
\includegraphics[clip, width=\hsize]{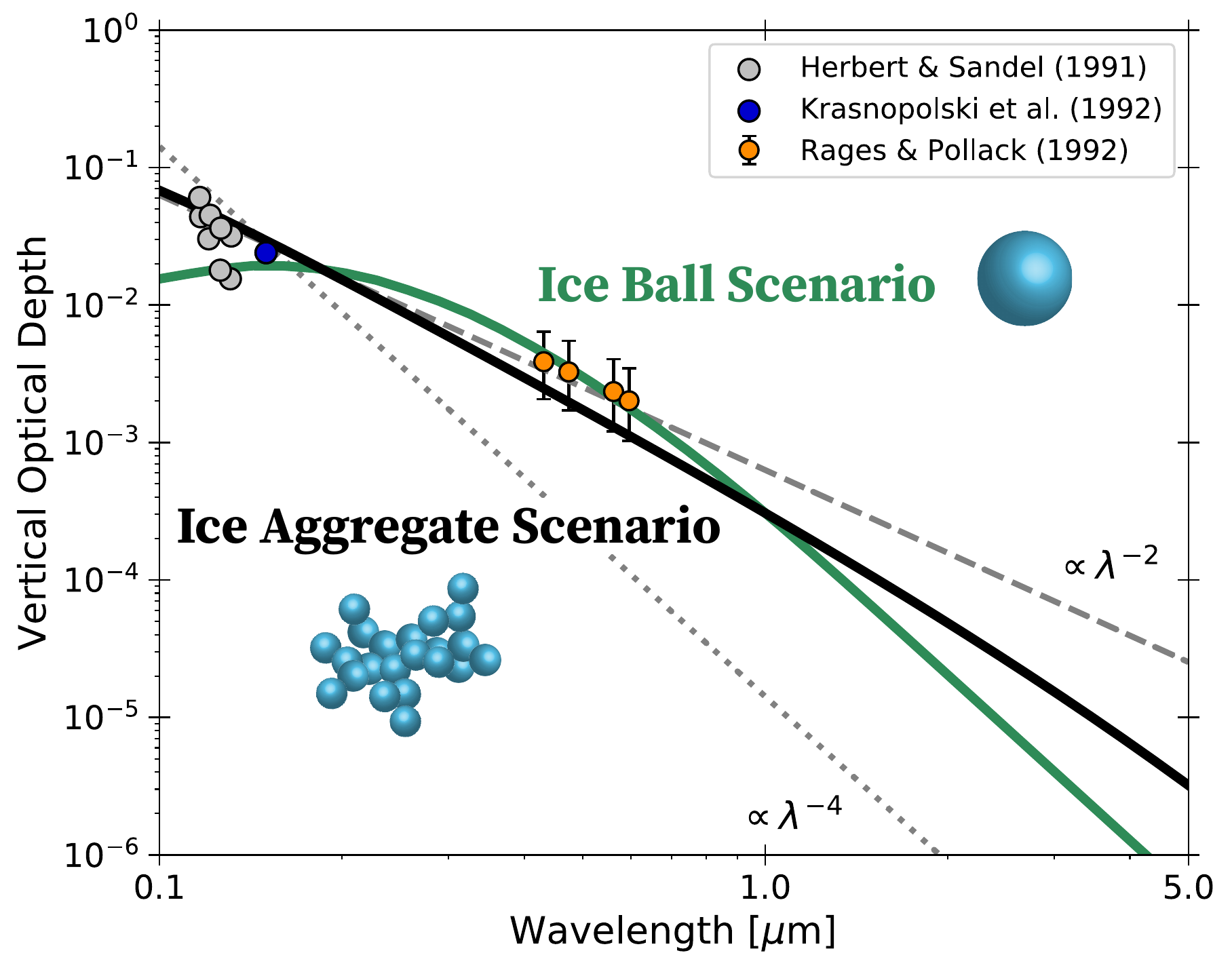}
\includegraphics[clip, width=\hsize]{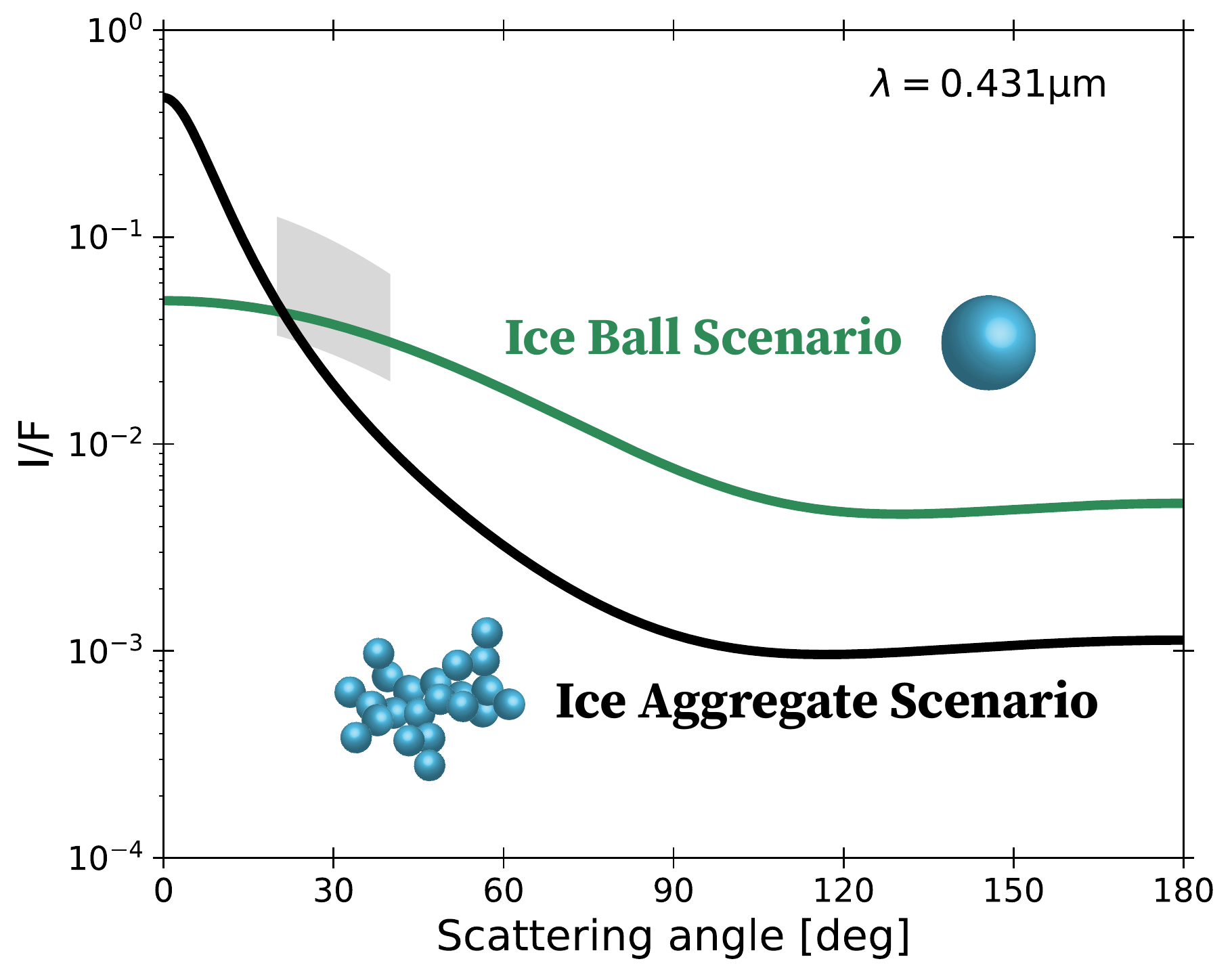}
\caption{
(Top) Vertical optical depth of the best models of ice balls (green line) and ice aggregates (black line).
The gray, blue, and orange symbols denote the optical depth reported by \citet{Herbert&Sandel91}, \citet{Krasbopolsky+92}, and \citet{Rages&Pollack92}, respectively.
The extinction optical depth is identical to the scattering one since we have assumed the imaginary refractive index of zero for C$_2$H$_4$ ice.
(Bottom) The visible $I/F$ for all scattering angles in other best models. The gray shaded region denotes the $I/F$ constrained in \citet{Rages&Pollack92}.
}
\label{fig:tau_summary}
\end{figure}

\ko{Our results suggest that ice condensation plays essential roles in haze formation on Triton.
However, it is still unclear which ice balls and ice aggregates are a more plausible solution.}
Figure \ref{fig:tau_summary} shows the optical depth in the best models for the ice ball and ice aggregate scenario.
We note that the extinction optical depth is identical to the scattering optical depth (i.e., conservative scattering) in Figure \ref{fig:tau_summary}, as we have assumed a zero imaginary refractive index.
Both ice ball and ice aggregate models reasonably explain the UV and visible optical depth reported by \citet{Krasbopolsky+92} and \citet{Rages&Pollack92}.

The largest difference between the ice ball and ice aggregate models appears at wavelengths shorter than $0.15~{\rm {\mu}m}$.
The optical depth is nearly invariant with wavelength at $\lambda<0.15~{\rm {\mu}m}$ in the ice ball scenario, while the optical depth is higher at a shorter wavelength in the ice aggregate scenario.
\citet{Herbert&Sandel91} reported that the optical depth might slightly increase with decreasing wavelength in ultraviolet observations of Voyager 2 (their Figure 7, see also gray dots in Figure \ref{fig:tau_summary}).
This might be more consistent with the ice aggregate scenario. 
Future observations on Triton need to constrain the haze optical depth in the far-UV range with more wavelength coverage.

The degree of forward scattering also helps to distinguish the two scenarios.
The bottom panel of Figure \ref{fig:tau_summary} shows the visible $I/F$ of the ice ball and ice aggregate scenarios for a whole range of scattering angles.
From the scattering phase functions, the ice aggregates induce much stronger forward scattering and much less backward scattering than the ice balls.
This is because the ice aggregates need to grow into sizes much larger than wavelength to exhibit a scattering opacity proportional to $\lambda^{-2}$ \citep{Berry&Percival86}.
Since the existing photometric observations are available only for a narrow range of phase angles ($140^{\circ}$--$160^{\circ}$, or scattering angles $20^{\circ}$--$40^{\circ}$), the scattering phase function cannot be well constrained.
Future observations, such as the NASA Ice Giants Mission\footnote{https://www.lpi.usra.edu/icegiants/} and the proposed NASA discovery-class mission TRIDENT \citep{Prockter+19,Mitchell+19}, would be greatly helpful to shed light on the morphological nature and formation processes of Triton hazes.

{ 
\section{Discussion}\label{sec:discussion2}

\subsection{Assessment of Ice Nucleation}\label{sec:nucleation}
In our icy haze scenarios, we have implicitly assumed that the nucleation of initial C$_2$H$_4$ ice embryos on photochemical haze can instantaneously occur.
However, the formation of icy hazes might be inhibited if the heterogeneous nucleation is inefficient.
Here we evaluate whether the heterogeneous nucleation of C$_2$H$_4$ ice is fast enough to form icy hazes.
We invoke the classical nucleation theory, reviewed in Appendix \ref{appendix:nucleation}.
We compare the nucleation timescale $J_{\rm het}^{-1}$, a time required to form one critical-sized embryo on a condensation nucleus, with the settling timescale of the nuclei $\tau_{\rm fall}=H/v_{\rm t}$.
If $\tau_{\rm fall}<J_{\rm het}^{-1}$, the condensation nucleus falls too fast to create the ice embryo on the nucleus surface, inhibiting the ice-coated haze formation.
On the other hand, if $\tau_{\rm fall}>J_{\rm het}^{-1}$, heterogeneous nucleation is fast enough to form icy hazes.
We evaluate the settling timescale assuming the material density of the condensation nuclei of $\rho_{\rm p}=1~{\rm g~{cm}^{-3}}$ and atmospheric density at $z=20~{\rm km}$.
Since the characteristic curvature radius is the monomer size, the ice condensation can occur if $J_{\rm het}^{-1}<\tau_{\rm fall}$ at a nuclei size smaller than $r_{\rm mon}$, which is $10~{\rm nm}$ in our fiducial simulations.

Whether nucleation and subsequent condensation of C$_2$H$_4$ ice can occur highly depends on the contact angle of C$_2$H$_4$ ice onto the haze particles, which is currently unknown.
Figure \ref{fig:Nucl} compares the heterogeneous nucleation timescale with the settling timescale of condensation nuclei.
For the nucleation through direct vapor deposition (top panel), the nucleation timescale goes below the settling timescale at nuclei radii of $\sim10$, $5$, and $3~{\rm nm}$ for saturation ratios of $S=3$, $10$, and $30$, respectively, when the contact angle is $\theta_{\rm c}=8.1^{\circ}$ ($\mu=0.99$).
If the contact angle is large, the nucleation is possible only when the saturation ratio is very high; for example, the nucleation timescale goes below the settling timescale at ${\sim}20~{\rm nm}$ for $S=30$ when $\theta_{\rm c}=25.84^{\circ}$ ($\mu=0.9$). 
The nucleation through surface diffusion of already adsorbed molecules can render the nucleation efficient (bottom panel).
Since the saturation ratio can be as high as $3$--$10$ in our fiducial simulations (Figure \ref{fig:ice_ball_revise}), the C$_2$H$_4$ nucleation would occur for ${\sim}10~{\rm nm}$ sized monomers as long as the contact angle is smaller than ${\sim}20^{\circ}$.

Several factors potentially facilitate the ice nucleation.
For example, surface roughness on condensation nuclei can enhance the nucleation rate \citep{Mahata&Alofs75}.
Charged condensation nuclei can also significantly enhance the heterogeneous nucleation, as discussed in the context of mesospheric cloud formation on Earth \citep{Gumbel&Megner09,Megner&Gumpel09}.
\citet{Wong+17} suggested that the C$_2$H$_4$ vapor pressure at cold temperatures is orders of magnitude lower than the extrapolation from experimental vapor pressure, which may lead to high supersaturation for C$_2$H$_4$.
The New Horizons observations showed a local minimum in the C$_2$H$_4$ vertical profile in Pluto's atmosphere, which is probably caused by the C$_2$H$_4$ ice condensation \citep{Wong+17,Young+18}. This suggests that C$_2$H$_4$ ice condensation is also likely to happen in the even colder Triton atmosphere. 

\begin{figure}[t]
\centering
\includegraphics[clip, width=\hsize]{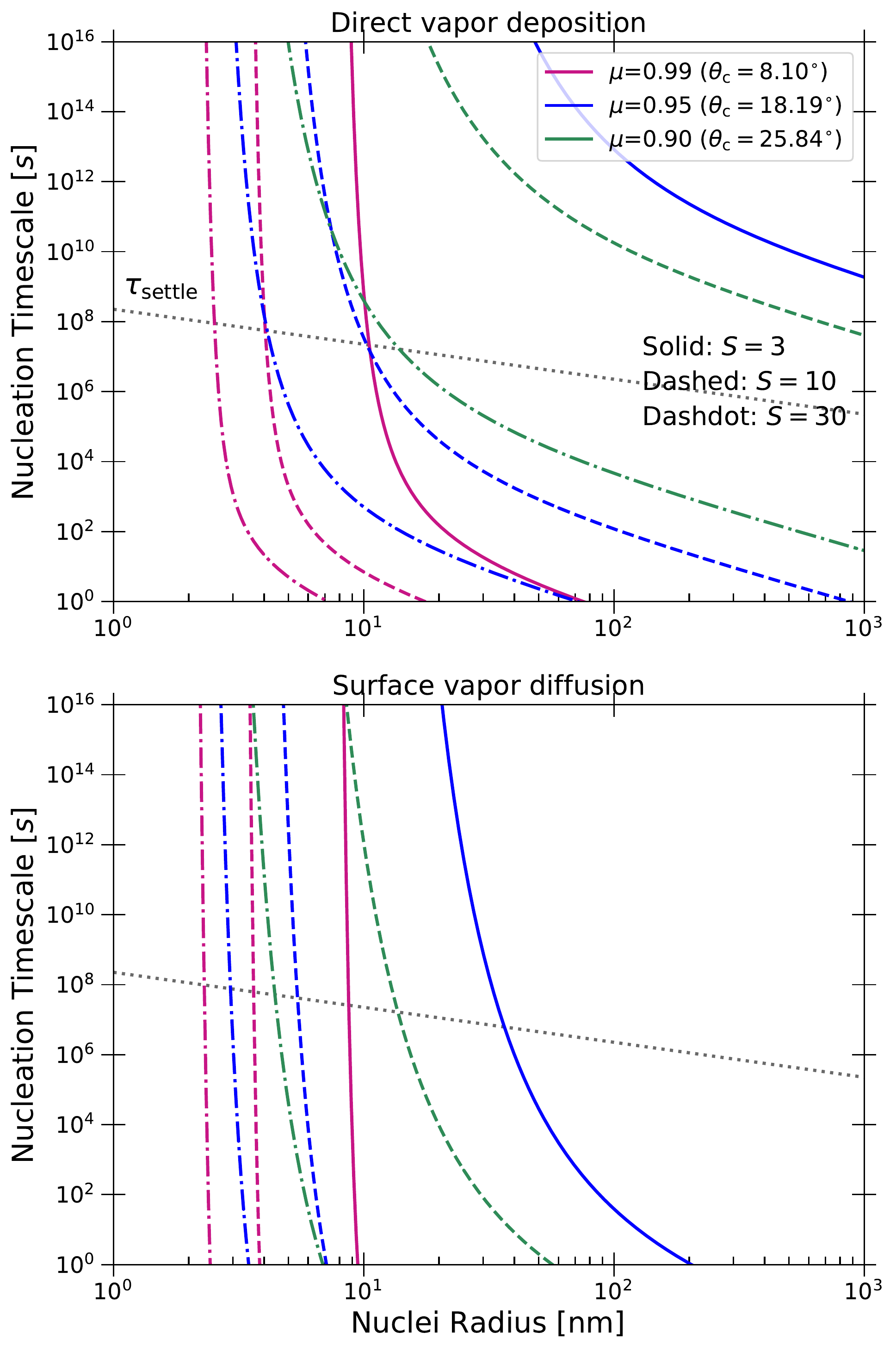}
\caption{
  Timescales of heterogeneous nucleation of C$_2$H$_4$ ice $J_{\rm het}^{-1}$ as a function of the radius of the condensation nuclei. The red, blue, and green lines show the timescale for the contact parameter of $\mu=0.99\ (\theta_{\rm c}=8.10^{\circ})$, $0.95\ (\theta_{\rm c}=18.19^{\circ})$, and $0.90\ (\theta_{\rm c}=25.84^{\circ})$, respectively. The solid, dashed, and dashdot lines show the timescale for $S=3$, $10$, and $30$, respectively.
The top and bottom panels show the timescales for heterogeneous nucleation through direct vapor deposition and surface diffusion of adsorbed molecules (see Appendix \ref{appendix:nucleation}).
The gray dotted lines denote the settling timescale of the condensation nuclei at $z=20~{\rm km}$.
Temperature is set to $T=50~{\rm K}$.
}
\label{fig:Nucl}
\end{figure}
\begin{figure*}[t]
\centering
\includegraphics[clip, width=\hsize]{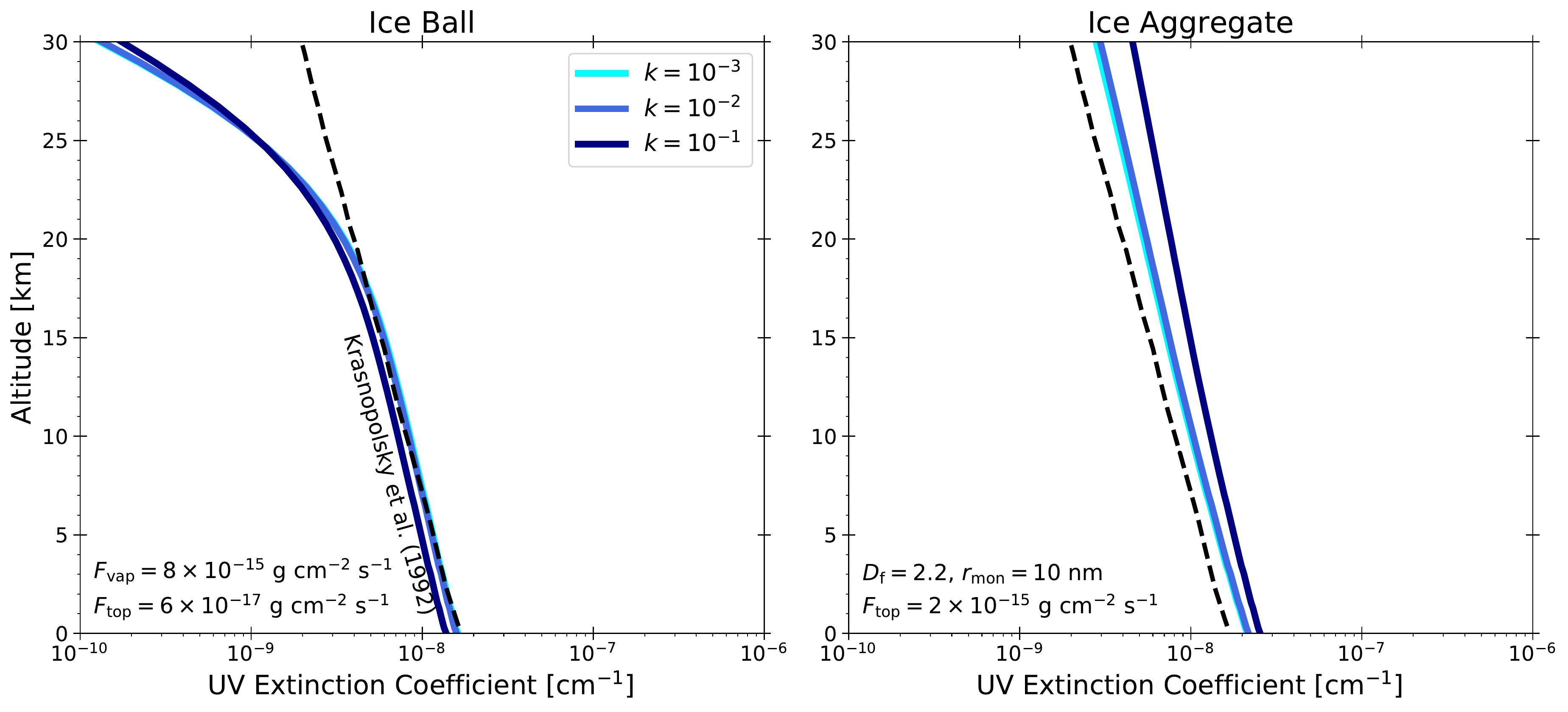}
\caption{
  UV extinction coefficient of icy hazes for non-zero imaginary refractive indices. The cyan, blue, and navy lines show the coefficient for $k={10}^{-3}$, ${10}^{-2}$, and ${10}^{-1}$, respectively. We set model parameters of our best ice ball and ice aggregate models (see Figure \ref{fig:iceball}) in the left and right panels, respectively.
}
\label{fig:test_k}
\end{figure*}

\subsection{{ Sensitivity} on Optical Constants of Icy Hazes}\label{sec:test_k}
We have assumed that the imaginary refractive index of icy hazes is zero at $\lambda=0.15~{\rm {\mu}m}$.
However, hydrocarbon ices themselves may cause substantial absorption in UV wavelengths.
While optical constants of ices in UV wavelengths have been available only for limited species \citep[e.g.,][]{Schmitt+98,Hendrix+13}, some of them exhibit strong absorption.
For example, H$_2$O ice has the imaginary refractive index of $k\sim0.3$ at $\lambda=0.15~{\rm {\mu}m}$ \citep{Warren&Brandt08}, and NH$_3$ ice has an index of $k\sim0.6$ \citep{Martonchik+84}.
Note that the index of Titan's tholin is $k=0.25$ \citep{Khare+84}.
On the other hand, some substances are transparent at the relevant UV wavelength.
For example, the imaginary refractive index of CH$_4$ ice steeply decreases to $k\sim{10}^{-5}$ from $\lambda\ga0.13~{\rm {\mu}m}$ \citep{Martonchik&Orton94}, and the index of CO$_2$ ice also decreases to $k\sim{10}^{-6}$ at $\lambda\ga0.12~{\rm {\mu}m}$ \citep{Warren86}.
{
The ice ball scenario hardly depends on the detail of the UV optical constants because the haze opacity falls into the geometric limit in UV wavelength.
On the other hand, as argued in Section \ref{sec:UV},
hazes need to be composed of almost nonabsorbing materials to simultaneously explain the UV and the visible observations when the Triton hazes are ice aggregates.}
Thus, it is vital to assess the effects of the nonzero imaginary refractive index on our results.

We calculate the UV extinction coefficient of icy hazes by varying an imaginary refractive index.
Figure \ref{fig:test_k} shows the UV extinction coefficient of haze particles for various imaginary refractive indices.
The UV extinction coefficient is almost invariant with the assumed imaginary refractive index for $k\la{10}^{-2}$, implying that the absorption is unimportant as long as $k\la{10}^{-2}$.
For $k=0.1$, the absorption starts to enhance the extinction coefficient for the ice aggregate case while slightly reduces the extinction for the ice ball case.
The latter behavior originates from the absorption that thwarts the interference of scattered light \citep{Bohren&Huffman83}. 
{ These results indicate that the contribution of absorption is negligible for $k\la0.1$ in our model set.}

{

{ Thus, our ice aggregate scenario would hold as long as $k\la0.1$.}
The condition is satisfied if $k=0.02$, which is inferred from the refractive index of CH$_4$ ice in \citet[][]{Krasbopolsky+92}.
On the other hand, \citet{Lavvas+20} recently estimated $k\sim0.7$ for C$_2$H$_4$ ices from the gas opacity of C$_2$H$_4$.
If this is true, the ice aggregates still cause absorption in UV wavelengths and hardly remedy the model--data discrepancy.
The simulation of \citet{Lavvas+20} seems to face the same difficulty for explaining both UV and visible observations simultaneously (see their Figure 8).
Future laboratory studies could validate the estimated optical properties of C$_2$H$_4$ ice in \citet{Lavvas+20}. 
If the high imaginary index of C$_2$H$_4$ ice is true, this implies that the ice ball scenario is a better solution for explaining observations of Triton hazes. 
}


\subsection{Why are Triton and Pluto Hazes different?}
Triton's N$_2$-CH$_4$-CO atmosphere is similar to that of Pluto; nevertheless, our results suggest that the properties of Triton hazes are quite different from those of Pluto hazes.
According to our results, Triton hazes are predominantly composed of hydrocarbon ices, while Pluto hazes are suggested to be similar to Titan hazes \citep{Gladstone+16,Gao+17,Cheng+17}, although there are no direct observational constraints on the Pluto haze compositions. 
{ We also note that a recent study of \citet{Lavvas+20} suggested that Pluto hazes may be composed of hydrocarbon ices, such as C$_4$H$_2$.
Nonetheless, it would be important to understand the cause in terms of their composition and formation processes if the Triton and Pluto hazes were indeed different.
}

One straightforward explanation of the difference is the temperature structure.
Triton's lower atmosphere is very cold ($\sim40$--$60~{\rm K}$), while the Pluto's lower atmosphere is relatively hot ($\sim 60$--$100~{\rm K}$) \citep[e.g.,][]{Strobel&Zhu17}.
Thus, the ice condensation onto haze particles can be inhibited at Pluto's lower hot atmosphere, explaining why Pluto hazes look different from Triton hazes. 
We note that the C2 hydrocarbons might condense or stick onto the haze particles in the cold upper atmosphere of Pluto \citep[e.g.,][]{Wong+17,Luspay-Kuti+17}.

The CH$_4$ abundance in the upper atmosphere might also cause the difference of Triton and Pluto haze properties.
The Voyager 2 observation found that CH$_4$ abundance steeply decreases with increasing altitudes on Triton \citep{Herbert&Sandel91}, while the abundance rather increases with increasing altitudes on Pluto as revealed by the New Horizons spacecraft \citep{Gladstone+16,Young+18}.
The New Horizons observation also found that high order hydrocarbons, such as C$_2$H$_2$, C$_2$H$_4$, and C$_2$H$_6$, are abundantly present in the upper atmosphere of Pluto \citep{Gladstone+16,Young+18}.
Since the photochemistry of hydrocarbons eventually yields a photochemical haze, as known from Titan, Pluto seems to be more favored to form Titan-like hazes than Triton.

Water delivery by interplanetary dust particles (IDPs) might be an alternative factor that causes the different haze formation processes.
In the outer solar system relevant to Triton and Pluto, IDPs are mostly coming from the Edgeworth-Kuiper Belt and Oort Cloud comets \citep{Poppe16}. 
Based on dust dynamics simulations, \citet{Poppe&Horanyi18} suggested that meteoroidal water influx on Triton is 2 orders of magnitude higher than that on Pluto because of strong gravitational acceleration and focusing by Neptune's gravity field.
The estimated H$_2$O mass flux is $1.8\times{10}^{-17}~{\rm g~{cm}^{-2}~s^{-1}}$ \citep{Poppe&Horanyi18}, which is about 2 orders of magnitude lower than the total haze mass flux suggested in this study.
However, the deposited water can influence the photochemistry of hydrocarbons; for example, photolysis of H$_2$O produces OH radicals that eventually form CO through reactions with carbon-based molecules \citep{Krasnopolsky12,Moses&Poppe17}.
It would be interesting to study how the IDPs' water delivery may affect hydrocarbon photochemistry and subsequent haze formation.



}



\section{Summary}\label{sec:summary}
In this study, we have presented the first microphysical model of haze formation on Triton.
Our model simulates the evolution of both size and porosity distributions of haze particles in a self-consistent manner.
We have compared the model results with the observed UV extinction coefficient and visible scattered-light intensity from Voyager 2.
We have shown that ice-free hazes, often assumed for Titan and Pluto hazes, cannot explain the Triton observations.
{Our results support the idea that Triton hazes are predominantly composed of hydrocarbon ices, which has been inferred from the Triton's cold environment but not assessed in detail.}
We have proposed two possible models of haze formation with ice condensation, namely ice ball and ice aggregate scenarios, that can successfully explain the existing observations of Triton hazes. 
Our findings are summarized as follows.

\begin{enumerate}
\item Haze particles {  can} grow into fractal aggregates even in the Triton's tenuous atmosphere (Section \ref{sec:vertical_icefree}).
The aggregates can grow to the mass-equivalent sphere radius of $0.2$--$1~{\rm \mu m}$, while the spheres can grow to only small sizes of $0.03$--$0.06~{\rm \mu m}$.
Due to collisional growth, the fractal dimension of fractal aggregates is $D_{\rm f}=1.8$--$2.2$, varying with the particle mass and altitude.
The mass-dominating aggregates have the fractal dimension of $D_{\rm f}\approx1.9$.
{The obtained $D_{\rm f}$ is in agreement with the outcome of cluster-cluster aggregation \citep[e.g.,][]{Meakin91} and} similar to the fractal dimension of Titan hazes.

\item Haze vertical profiles substantially vary with downward mass flux, monomer sizes, and charge density (Section \ref{sec:vertical_icefree}).
In general, higher mass flux results in larger particle sizes and higher mass density.
By contrast, larger monomers tend to produce smaller aggregates with lower mass densities.
The high particle charge reduces the particle size but has minor impacts on the mass density.
The mass-dominating aggregates have $D_{\rm f}\approx1.9$ for almost all mass fluxes, monomer sizes, and particle charge in this study.

\item Ice-free hazes cannot explain both the UV extinction coefficient and visible scattered-light intensity simultaneously (Section \ref{sec:UV}).
Both the aggregate and sphere models match the UV extinction coefficient if the haze mass flux is $F_{\rm top}{\sim}3\times{10}^{-15}~{\rm g~{cm}^{-2}~s^{-1}}$, whereas the models could match the visible $I/F$ only when the mass flux is $F_{\rm top}{\sim}3\times{10}^{-14}~{\rm g~{cm}^{-2}~s^{-1}}$.

\item The discrepancy is attributed to the wavelength dependence of opacity for absorbing hazes (Section \ref{sec:UV}).
The UV extinction and visible scattering optical depth of Triton hazes are nearly proportional to $\lambda^{-2}$.
The ice-free spheres cannot explain this dependence because the particles are so small that they induce Rayleigh scattering.
The ice-free aggregates also fail to explain the wavelength dependence owing to low single-scattering albedo.
Different monomer sizes and charge densities cannot reconcile the discrepancy.

\item We suggest that condensation of hydrocarbon ices plays a vital role in haze formation on Triton (Section \ref{sec:discussion}).
For icy spherical hazes (ice ball scenario), the model could explain both UV extinction coefficient and visible $I/F$ for the downward ice-free mass flux of $F_{\rm top}=6\times{10}^{-17}~{\rm g~{cm}^{-2}~s^{-1}}$ {  when the column-integrated C$_2$H$_4$ production rate is $F_{\rm vap}=8\times{10}^{-15}~{\rm g~{cm}^{-2}~s^{-1}}$.}
For icy aggregates, the model could explain both UV and visible data for the mass flux of $F_{\rm top}=2\times{10}^{-15}~{\rm g~{cm}^{-2}~s^{-1}}$ when the fractal dimension is $D_{\rm f}\approx2.2$.
The required mass flux is comparable to the column-integrated production rate of condensable hydrocarbons, $4$--$8\times{10}^{-15}~{\rm g~{cm}^{-2}~s^{-1}}$ \citep[][]{Strobel+90,Strobel&Summers95}. 

\item Future observations of the UV optical depth with greater wavelength coverage and scattering phase function with more phase angles would distinguish the ice ball and ice aggregate scenarios.
The optical depth of ice aggregates increases with decreasing the wavelength at $\lambda<0.15~{\rm \mu m}$, while the optical depth of the ice balls is invariant at this wavelength range. 
The ice aggregates are slightly more consistent with the UV solar occultation observations of Voyager 2.
The ice aggregates also cause forward scattering stronger than the ice balls do.
These observational signatures would help to shed light on the nature of haze formation on Triton for future observations, such as the NASA Ice Giants Mission and TRIDENT.

\end{enumerate}


{
During the revision of this paper, a contemporaneous study of \citet{Lavvas+20} also provided a microphysical model of Triton hazes, though their main focus was Pluto hazes.
\citet{Lavvas+20} considered aggregate hazes composed of multiple hydrocarbon ices. 
Their findings are broadly consistent with our aggregate scenario.
For example, both our model and \citet{Lavvas+20} predicted that collisional aggregation takes place only at $z\la50~{\rm km}$ when the monomer radius is $\sim20~{\rm nm}$ (Figure \ref{fig:vertical_profiles}).
Our icy haze models also suggest the total haze mass flux of $2$--$8\times{10}^{-15}~{\rm g~{cm}^{-2}~s^{-1}}$, depending on the assumed particle morphology, which is in agreement with the mass flux of $\sim6\times{10}^{-15}~{\rm g~{cm}^{-2}~s^{-1}}$ suggested by \citet{Lavvas+20}.
We have suggested hydrocarbon ices as the predominant compositions of Triton hazes from the Titan tholin's inability to explain the observations, while \citet{Lavvas+20} obtained predominant haze composition of C$_2$H$_4$ ices from the output of the photochemical model.
Our study and \citet{Lavvas+20} faced the same difficulty in explaining both UV and visible observations if the aggregate is made of absorbing materials.
Future laboratory studies will be needed to measure the optical properties of C$_2$H$_4$ ices.
Alternatively, we suggest the ice ball scenario as a plausible solution if C$_2$H$_4$ ice is absorbing in UV.

}

{  
Our study highlights the importance of future laboratory studies applicable to extremely cold atmospheres.
As demonstrated in this study, hydrocarbon ices likely play essential roles in aerosol formation in the outer solar system. 
However, their optical properties are currently uncertain at visible-to-UV wavelengths \citep{Schmitt+98,Hendrix+13}, which limit our ability to interpret observations.
Laboratory studies of vapor pressure for extremely cold temperatures will also be warranted to better understand how aerosols grow and how the gaseous molecules are removed through condensation.
To better understand the microphysical processes of aerosol formation, it will be vital to know the desorption energy and contact angle of hydrocarbon ices on tholin from laboratory studies, which is currently available only for CH$_4$ and C$_2$H$_6$ \citep{Curtis+08,Rannou+19}.
These laboratory studies will be greatly helpful for interpreting the observations of the ongoing New Horizons mission and the future NASA Ice Giants Mission and TRIDENT.
}

Lastly, although we did not focus on atmospheric thermal structure in this study, hazes may play an important role in controlling the temperature structure on Triton.
\citet{Zhang+17} suggested that radiative cooling by hazes is a key to explaining the cold temperature on Pluto.
It would be interesting to include the feedback of haze radiative effects on Triton's temperature structure in future haze formation models. 
{  Since our results suggest that Triton hazes at the lower atmosphere are likely composed of hydrocarbon ices, the haze radiative feedback may have different effects as compared to that on Pluto, for which optical constants of the Titan tholin are often assumed.
On the other hand, since Triton hazes are likely ice-free in the hot upper atmosphere, they may act as coolants in the upper atmosphere, as suggested for Pluto.
}



\acknowledgments 
We thank Darrell Strobel for providing Triton's temperature structure. 
{We are also grateful to the anonymous referee for a number of suggestions that greatly improved the quality of this paper.}
This work is supported by JSPS KAKENHI Grant Nos. JP18J14557, JP18H05438, and JP19K03926. 
{ K.O. acknowledges support from the JSPS Overseas Research Fellowships.}
X. Z. is supported by NASA Solar System Workings Grant 80NSSC19K0791.


\appendix
\section{Analytical Absorption and Scattering Coefficients of Fractal Aggregates}\label{appendix:anal}

\ko{In this appendix, we present an analytical theory that predicts the absorption and scattering coefficients of aggregate hazes.
The theory is useful for understanding how observable quantities depend on our model parameters, namely the haze mass flux, monomer size, and charge density.
We also expect that the theory helps to quickly evaluate the haze parameters from remote-sensing observations before using a detailed microphysical model.
Our theory is based on the assumption that incident light is only scattered once by every monomer within an aggregated particle.
In particular, the single-scattering assumption holds for aggregates with $D_{\rm f}\le2$ constituted by monomers much smaller than the wavelength \citep[][]{Berry&Percival86,Tazaki&Tanaka18}.
}
Under this assumption, the absorption cross section is approximated by \citep[Section 5 of][]{Berry&Percival86}
\begin{equation}\label{eq:sigma_abs_agg}
    \sigma_{\rm abs,agg}=N_{\rm mon}\sigma_{\rm abs,mon},
\end{equation}
where $\sigma_{\rm abs,mon}$ is the absorption cross section of a monomer.
We note that Eq \eqref{eq:sigma_abs_agg} does not hold for aggregates with large refractive indices for which monomer--monomer interaction plays an important role \citep{Tazaki&Tanaka18}.
On the other hand, the scattering cross section of an aggregate $\sigma_{\rm sca,agg}$ is approximated by \citep[][]{Berry&Percival86}
\begin{equation}
\label{eq:Berry}
    \frac{\sigma_{\rm sca,agg}}{N_{\rm mon}\sigma_{\rm sca,mon}}\approx 
\left\{
\begin{array}{ll}
    {\displaystyle  \frac{2\cos{[(2-D_{\rm f})\pi/2]}}{(D_{\rm f}-1)(2-D_{\rm f})}\left( \frac{b\lambda}{4\pi r_{\rm mon}}\right)^{D_{\rm f}} } & \text{($D_{\rm f}<2$)}  \\[1.5ex]
      {\displaystyle \log{\left( \frac{16\pi^2r_{\rm mon}^2N_{\rm mon}}{b\lambda^2}\right)}\left( \frac{b\lambda}{4\pi r_{\rm mon}}\right)^{2} }& \text{($D_{\rm f}=2$)} \\[1.5ex]
      {\displaystyle \frac{2N_{\rm mon}^{1-2/D_{\rm f}}}{(D_{\rm f}-1)(2-D_{\rm f})}\left( \frac{b\lambda}{4\pi r_{\rm mon}}\right)^{2} } & \text{($D_{\rm f}>2$)} ,
         \end{array}
\right.
\end{equation}
where $b$ is a constant order of unity, $\sigma_{\rm sca,mon}$ is the scattering cross section of a monomer, and we have assumed the aggregates much larger than the wavelength ($r_{\rm agg}\gg \lambda/2\pi$).
It may be reasonable to assume that the monomers are much smaller than the wavelength. 
In that case, the absorption and scattering cross sections of a monomer are respectively approximated by \citep{Bohren&Huffman83,Kataoka+14}
\begin{equation}\label{eq:Rayleigh_abs}
    \sigma_{\rm abs,mon}\approx \pi r_{\rm mon}^{2} \times \frac{24nk}{(n^2-k^2+2)^2+(2nk)^2}\left( \frac{2\pi r_{\rm mon}}{\lambda}\right),
\end{equation}
and 
\begin{equation}\label{eq:Rayleigh}
    \sigma_{\rm sca,mon}\approx \pi r_{\rm mon}^{2} \times \frac{32}{27}\left( \frac{2\pi r_{\rm mon}}{\lambda}\right)^4 \left[ (n-1)^2+k^2\right],
\end{equation}
where $n$ and $k$ are the real and imaginary parts of the refractive index.
From Equations \eqref{eq:Berry} and \eqref{eq:Rayleigh}, the scattering cross section of an aggregate is proportional to $\lambda^{D_{\rm f}-4}$ for $D_{\rm f}<2$ and $\lambda^{-2}$ for $D_{\rm f}\geq2$.
Notably, the dependence for $D_{\rm f}\geq2$ is similar to the spectral behavior suggested by observations of Triton hazes.

In what follows, we derive the absorption and scattering coefficient for the aggregates with $D_{\rm f}=2$.
The aggregates with $D_{\rm f}=2$ have been suggested for Titan hazes \citep{Rannou+97}, assumed in many previous studies \citep[e.g.,][]{Rannou+03,Tomasko+08,Lavvas+10,Gao+17}, and in agreement with our simulated $D_{\rm f}$ (Section \ref{sec:vertical_icefree}).
Using Equations \ref{eq:sigma_abs_agg} and \eqref{eq:Rayleigh_abs}, the absorption coefficient is evaluated as
 \begin{equation}\label{eq:alpha_abs0}
    \alpha_{\rm abs}= \int \frac{\sigma_{\rm abs,agg}}{m}mn(m)dm =  \frac{3\pi}{2 \rho_{\rm 0}\lambda}\rho_{\rm haze} \frac{24nk}{(n^2-k^2+2)^2+(2nk)^2},
\end{equation}
where we have used $m=m_{\rm mon}N_{\rm mon}$ and $\rho_{\rm haze}\equiv \int mn(m)dm$.
Similarly, using Equations \eqref{eq:Berry} and \eqref{eq:Rayleigh}, the scattering coefficient is evaluated as
\begin{equation}\label{eq:alpha_sca0}
    {\displaystyle \alpha_{\rm sca}= \int \frac{\sigma_{\rm sca,agg}}{m}mn(m)dm    = \frac{8\pi^2br_{\rm mon}}{9\rho_{\rm 0}\lambda^{2}} \rho_{\rm haze}\left[ (n-1)^2+k^2\right] \frac{\int \log{\left( 16\pi^2r_{\rm agg}^2/{b\lambda^2}\right)} mn(m)dm}{\int mn(m)dm}}.
\end{equation}
The haze mass density $\rho_{\rm haze}$ can be evaluated from the mass conservation (Equation \ref{eq:rho_haze}).
As a good approximation, the settling velocity of aggregates with $D_{\rm f}=2$ can be estimated as
\begin{equation}\label{eq:vt_appro}
    v_{\rm t}\approx \frac{\rho_{\rm p}gr_{\rm mon}}{\rho_{\rm g}C_{\rm s}}.
\end{equation}
The mass-averaged settling velocity $\overline{v}_{\rm t}$ is identical to Equation \eqref{eq:vt_appro} for any size distributions, as the velocity is independent of the aggregate size.
Thus, Equation \eqref{eq:rho_haze} yields the steady-state haze mass density of
\begin{equation}\label{eq:rho_haze_appendix}
    \rho_{\rm haze}\approx\frac{\rho_{\rm g}C_{\rm s}F_{\rm top}}{\rho_{\rm 0}gr_{\rm mon}}
\end{equation}
Combining Equations \eqref{eq:alpha_abs0}, \eqref{eq:alpha_sca0}, and \eqref{eq:rho_haze_appendix}, we finally achieve the absorption and scattering coefficient of the aggregates as
\begin{equation}\label{eq:alpha_abs_Df2}
    \alpha_{\rm abs}= \frac{3\pi \rho_{\rm g}C_{\rm s}F_{\rm top}}{2 \rho_{\rm 0}^2gr_{\rm mon}\lambda} \frac{24nk}{(n^2-k^2+2)^2+(2nk)^2},
\end{equation}
\begin{equation}\label{eq:alpha_sca_Df2}
    \alpha_{\rm sca}=\frac{8\pi^2\rho_{\rm g}C_{\rm s}F_{\rm top}}{9\rho_{\rm 0}^2g\lambda^{2}} \left[ (n-1)^2+k^2\right] \frac{\int \log{\left( 16\pi^2r_{\rm agg}^2/{b\lambda^2}\right)} mn(m)dm}{\int mn(m)dm}
\end{equation}
Equations \eqref{eq:alpha_abs_Df2} and \eqref{eq:alpha_sca_Df2} clarify how the absorption and scattering coefficients depend on haze properties.
For example, both the absorption and scattering coefficients are nearly independent of the aggregate sizes, except for logarithmic dependence in $\alpha_{\rm sca}$.
The absorption coefficient is inversely proportional to the monomer size, while the scattering coefficient does not explicitly depend on $r_{\rm mon}$. 
Both the absorption and scattering coefficients are proportional to the mass flux $F_{\rm top}$.
These parameter dependences would be useful to understand how we can infer the haze properties from observations.

We note that the above arguments for the scattering coefficient are violated when the aggregate is much smaller than the wavelength ($r_{\rm agg}\ll \lambda/2\pi$). 
In this case, the scattering cross section is approximated by \citep{Berry&Percival86}
\begin{equation}\label{eq:sigma_agg_Rayleigh}
    \sigma_{\rm sca,agg}\approx N_{\rm mon}^2\sigma_{\rm sca,mon}.
\end{equation}
Combining Equations \eqref{eq:Rayleigh}, \eqref{eq:alpha_sca0},  \eqref{eq:rho_haze_appendix}, and \eqref{eq:sigma_agg_Rayleigh}, we achieve the scattering coefficient of
\begin{equation}\label{eq:alpha_sca_Df2_Rayleigh}
    \alpha_{\rm sca}=\frac{128\pi^4\rho_{\rm g}C_{\rm s}F_{\rm top}}{9\rho_{\rm 0}^2g\lambda^{4}} \left[ (n-1)^2+k^2\right] \frac{\int r_{\rm agg}^2 mn(m)dm}{\int mn(m)dm}.
\end{equation}
Thus, for small aggregates, the scattering coefficient turns out to strongly depend on the aggregate sizes.

\begin{figure*}[t]
\centering
\includegraphics[clip, width=\hsize]{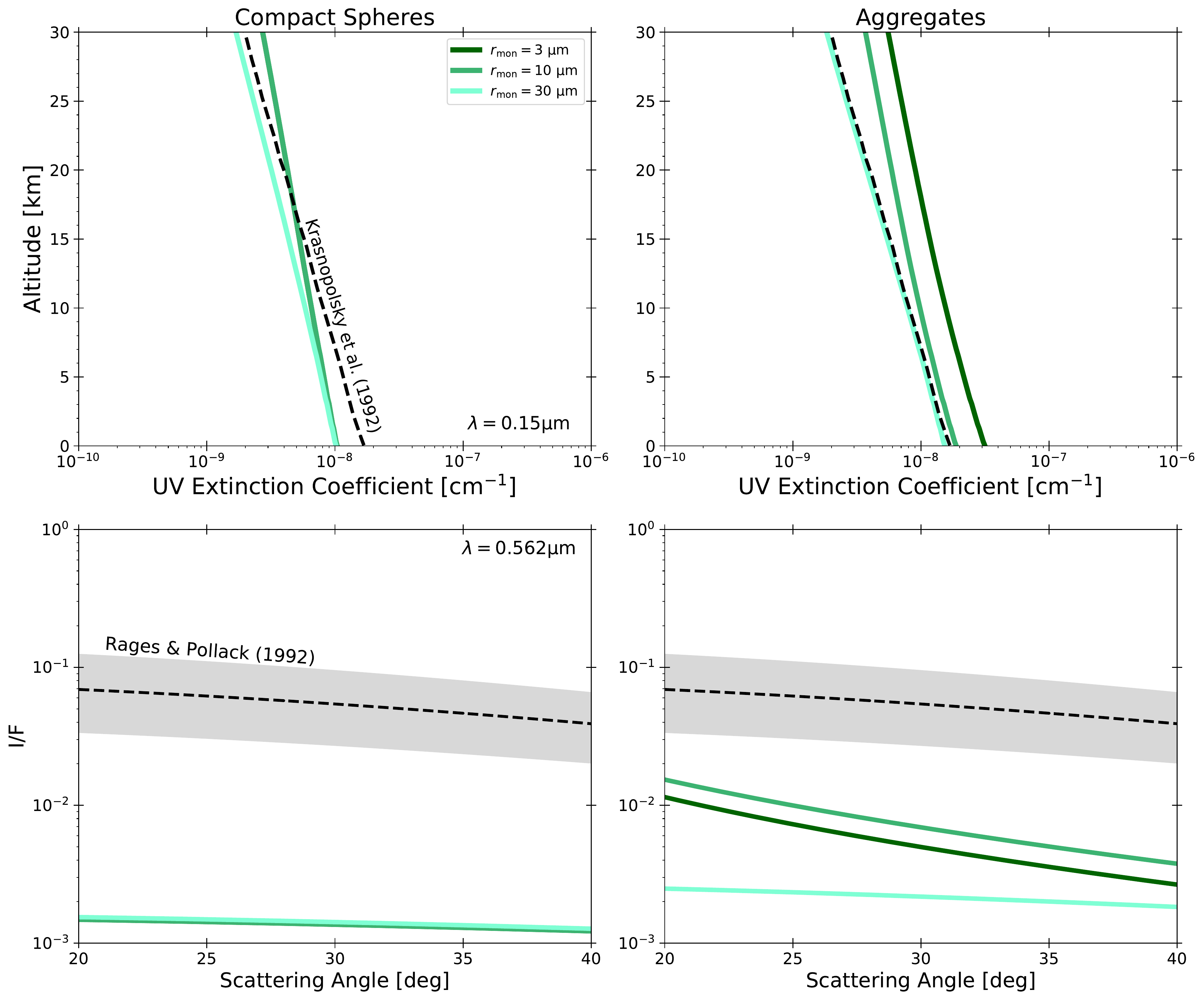}
\caption{Same as Figure \ref{fig:comparison_Flux}, but for different monomer sizes. The haze mass flux and particle charge are fixed to $F_{\rm top}=3\times{10}^{-15}~{\rm g~{cm}^{-2}~s^{-1}}$ and $q_{\rm e}=0$.}
\label{fig:comparison_rmon}
\end{figure*}
\begin{figure*}[t]
\centering
\includegraphics[clip, width=\hsize]{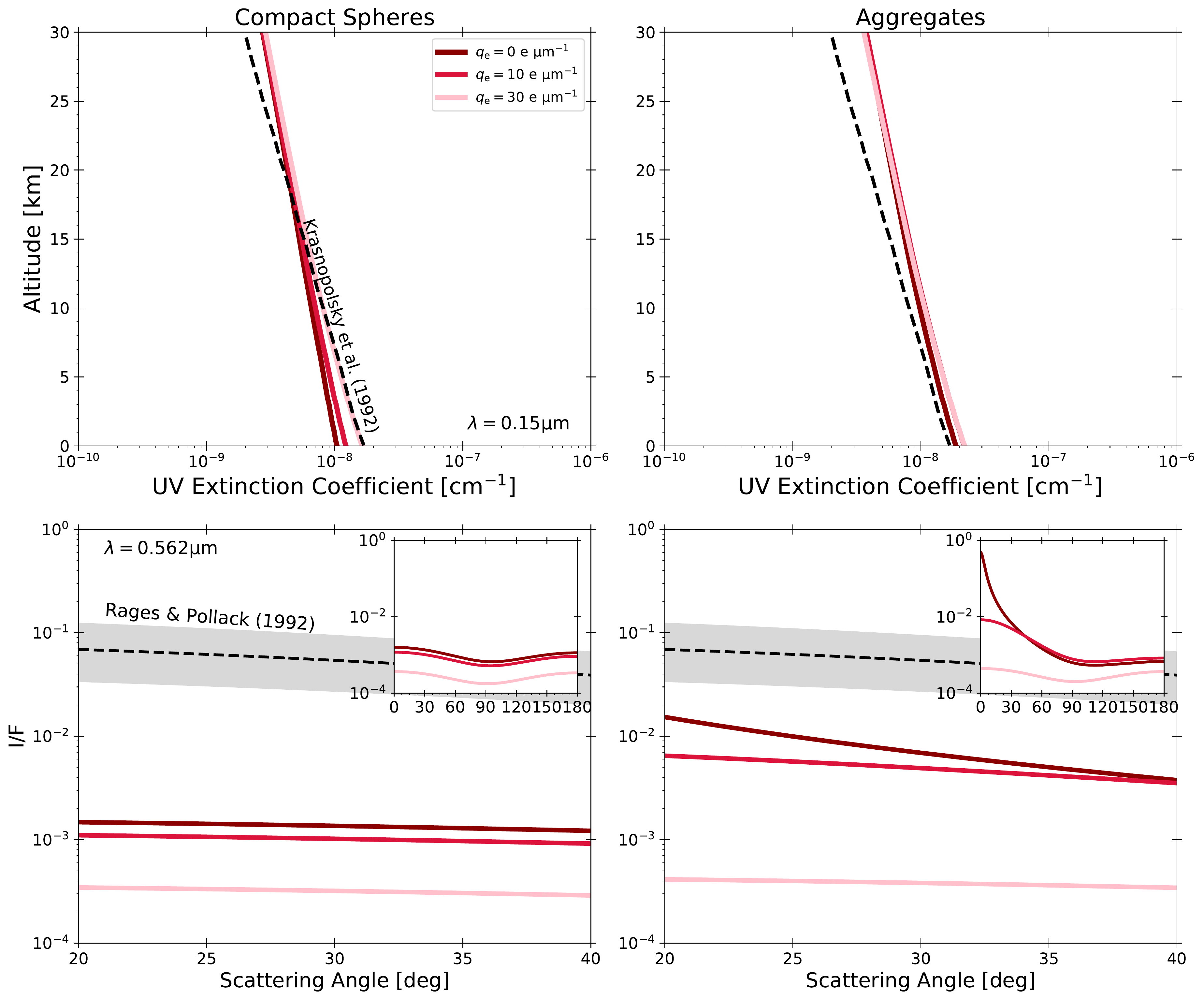}
\caption{Same as Figure \ref{fig:comparison_Flux}, but for different particle charge density. The haze mass flux and monomer size are fixed to $F_{\rm top}=3\times{10}^{-15}~{\rm g~{cm}^{-2}~s^{-1}}$ and $r_{\rm mon}=10~{\rm nm}$.
The subfigures in bottom panels show the $I/F$ for a whole range of scattering angles.
}
\label{fig:comparison_qe}
\end{figure*}




\subsection{Monomer Size Dependence}\label{appendix:monomer}
\ko{In this appendix, we discuss the effects of the monomer size on the observations of hazes. }
The top panels of Figure \ref{fig:comparison_rmon} show the UV extinction coefficient for different monomer sizes, assuming the refractive index of the Titan tholin \citep{Khare+84}.
The mass flux and particle charge are fixed to $F_{\rm top}=3\times{10}^{-15}~{\rm g~{cm}^{-2}~s^{-1}}$ and $q_{\rm e}=0$, respectively.
We note that the UV extinction coefficient is roughly the same as the absorption coefficient for the Titan tholin.
For the spheres, both the UV extinction coefficient and visible $I/F$ are insensitive to the monomer size because the vertical distributions are insensitive to the monomer size (see Figure \ref{fig:vertical_profiles}).
For the aggregates, smaller monomer sizes lead to a higher extinction coefficient because smaller monomers yield higher haze mass density.
This trend is in agreement with our analytical estimate of the absorption coefficient (Equation \ref{eq:sigma_abs_agg}). 
Since the absorption coefficient is proportional to $F_{\rm top}/r_{\rm mon}$, in general, it is hard to constrain the haze mass flux $F_{\rm top}$ from the extinction coefficient without knowledge of monomer sizes $r_{\rm mon}$, unless the single-scattering albedo is very high.

The visible $I/F$ is insensitive to the monomer size when the monomers are small (say, $r_{\rm mon}<30~{\rm nm}$).
The bottom panels of Figure \ref{fig:comparison_rmon} show that the visible $I/F$ for the spheres is nearly invariant with the monomer sizes for the same reason as the UV extinction coefficient.
The $I/F$ for the aggregates are also insensitive to the monomer size for $r_{\rm mon}=3$ and $10~{\rm nm}$.
This is because the small monomer size yields a low scattering opacity that largely cancels out the high mass density caused by small monomers.
This trend is also in agreement with our analytical estimate of the scattering coefficient (Equation \ref{eq:alpha_sca_Df2}).
On the other hand, the larger monomer size of $r_{\rm mon}=30~{\rm nm}$ yields an $I/F$ lower than those for $r_{\rm mon}=3$ and $10~{\rm nm}$.
The sharp drop of $I/F$ is attributed to the small sizes of aggregates that induce Rayleigh scattering.
Thus, for $r_{\rm mon}\ga30~{\rm nm}$, the $I/F$ is lower at larger $r_{\rm mon}$ that yield smaller aggregates (see also Equation \ref{eq:alpha_sca_Df2_Rayleigh}).


\subsection{Particle Charge Dependence}\label{appendix:charge}
\ko{In this appendix, we discuss the effects of the particle charge on the observations.}
The top panels of Figure \ref{fig:comparison_qe} show the UV extinction coefficient for different charge densities, assuming the refractive index of the Titan tholin.
The mass flux and monomer size are fixed to $F_{\rm top}=3\times{10}^{-15}~{\rm g~{cm}^{-2}~s^{-1}}$ and $r_{\rm mon}=10~{\rm nm}$, respectively.
Despite the effects on particle sizes, the extinction coefficient is insensitive to the particle charge.
This is in agreement with our analytical estimate of the absorption coefficient (Equation \ref{eq:alpha_abs_Df2}).
Since the particle charge only affects the aggregate size and does not affect the mass density, the extinction coefficient is insensitive to the particle charge unless the particle single-scattering albedo is very high.
This also explains why the extinction coefficient simulated by \citet{Gao+17} for the Pluto haze is insensitive to the particle charge (their Figure 5).

The particle charge appreciably affects the scattered-light intensity $I/F$, especially at low scattering angles.
The bottom panels of Figure \ref{fig:comparison_qe} show how the $I/F$ depends on the particle charge.
The higher the charge density, the weaker $I/F$ at small scattered angles.
This is because a higher charge density leads to smaller aggregate sizes, resulting in suppressing the forward scattering.
On the other hand, the $I/F$ at the scattering angles probed by Voyager 2 ($\theta=20^{\circ}$--$40^{\circ}$) is insensitive to the particle charge.
This is because the scattering coefficient is insensitive to the aggregate size (Equation \eqref{eq:alpha_sca_Df2}).
The particle charge eventually decreases the entire $I/F$ when the aggregates are smaller than the wavelength and fall into the Rayleigh regime (Equation \ref{eq:alpha_sca_Df2_Rayleigh}).

{ 
\section{Nucleation Theory}\label{appendix:nucleation}
In this appendix, we review the classical nucleation theory.
The formation of an initial condensate particle can be classified into either homogeneous or heterogeneous nucleation.
The former is the particle formation through the aggregation of gaseous molecules without external surfaces, while the latter is the particle formation onto already existing surfaces.
Atmospheric aerosols, such as photochemical hazes, can serve as condensation nuclei that provide the external surface for heterogeneous nucleation.
The nucleation theory provides the number of newly formed particles for a given time.
We refer readers to \citet[][]{Moses+92,pruppacher&Klett97,Seinfeld&Pandis06} for elaborate descriptions of the nucleation theory.

Generally speaking, the nucleation theory counts how many embryos---aggregates of molecules---become energetically stable per unit of time.
The actual energy of embryo formation depends on its complex structure. 
In practice, the classical nucleation theory assumes that the embryo has a spherical shape and bulk physical properties, such as material density and surface energy.
Under these assumptions, the formation energy for an embryo of i molecules (i-mer) is given by \citep[e.g.,][]{pruppacher&Klett97}
\begin{equation}\label{eq:F_i}
    \Delta F_{\rm i}=4\pi a_{\rm i}^2 \sigma - \frac{4\pi a_{\rm i}^3\rho_{\rm 0}}{3m_{\rm v}}k_{\rm B}T\ln{S},
\end{equation}
where $a_{\rm i}$ is the i-mer radius, $\sigma$ is the surface energy, $\rho_{\rm 0}$ is the material density, and $m_{\rm v}$ is the mass of a molecule. 
The first term stands for the energy required to form a new surface, while the second term stands for the decrease of chemical potential from gas to solid (or liquid) phases.
The equilibrium number density of i-mers $n_{\rm i}$ is associated with $\Delta F_{\rm i}$ as \citep[e.g.,][]{Tanaka+11}
\begin{equation}\label{eq:n_i}
    n_{\rm i}=\frac{\rho_{\rm v}}{m_{\rm v}}\exp{\left( -\frac{\Delta F_{\rm i}}{k_{\rm B}T}\right)},
\end{equation}
where $\rho_{\rm v}$ is the vapor mass density.
One can find that the formation energy of Equation \eqref{eq:F_i} has a single maximum at a certain i-mer radius, which is called the critical radius $a_{\rm g}$ and given by
\begin{equation}\label{eq:a_g}
    a_{\rm g}=\frac{2m_{\rm v}\sigma}{\rho_{\rm 0}k_{\rm B}T\ln{S}}.
\end{equation}
The corresponding formation energy of a critical-sized embryo is given by
\begin{equation}
    \Delta F_{\rm g}=\frac{4}{3}\pi a_{\rm g}^2\sigma.
\end{equation}
The embryo with a size of $a_{\rm i}>a_{\rm g}$ can stably exist as the formation energy decreases with increasing the embryo size.
Assuming that molecule--embryo collisions mainly drive the embryo growth, one can estimate the nucleation rate as the rate at which critical-sized embryos collect single molecules from the vapor.
For the homogeneous nucleation, the nucleation rate per unit volume ($J_{\rm hom}~{\rm [m^{-3}~s^{-1}]}$) is given by
\begin{equation}\label{eq:J_hom}
    J_{\rm hom}=4\pi a_{\rm g}^2 \Phi_{\rm v} Z\frac{\rho_{\rm v}}{m_{\rm v}}\exp{\left( -\frac{\Delta F_{\rm g}}{k_{\rm B}T}\right)},
\end{equation}
where $\Phi_{\rm v}$ is the number flux of vapor molecules, given by
\begin{equation}
    \Phi_{\rm v}=\frac{\rho_{\rm v}}{4m_{\rm v}}\sqrt{\frac{8k_{\rm B}T}{\pi m_{\rm v}}}=\frac{P_{\rm v}}{\sqrt{2\pi m_{\rm v}k_{\rm B}T}},
\end{equation}
and $Z$ is the so-called Zeldovich factor that accounts for the deviation of the embryo size distribution from the equilibrium distribution at $a_{\rm i}=a_{\rm g}$ (i.e., Equation \ref{eq:n_i}), given by
\begin{equation}\label{eq:Zeldovich}
    Z=\sqrt{ \frac{\Delta F_{\rm g}}{3\pi k_{\rm B}Tg_{\rm *}^2 }},
\end{equation}
where $g_{\rm *}$ is the number of molecules in a critical-sized embryo.
The Zeldovich factor typically has a value of $\sim0.1$.
As seen in Equation \eqref{eq:J_hom}, the nucleation rate is mostly controlled by the formation energy of the critical-sized embryo $\Delta F_{\rm g}$ because of the exponential nature.

It has been well known that nucleation onto external surfaces, i.e., heterogeneous nucleation, is usually much more efficient than homogeneous nucleation.
This is because the external surface can reduce the formation energy of an embryo.
A spherical cap on the surface can represent the shape of the embryo on an insoluble condensation nucleus. 
The critical radius of the spherical cap is the same as that for homogeneous nucleation (Equation \ref{eq:a_g}), but its formation energy is reduced from $\Delta F_{\rm g}$ by a factor of $f(\mu,x)$, which is called the shape factor and given by \citep{Frecher58,Fletcher59,Moses+92,pruppacher&Klett97}
\begin{equation}
    2f(\mu,x)=1+\left( \frac{1-\mu x}{\phi}\right)^3+x^3(2-3f_{\rm 0}+f_{\rm 0}^3)+3\mu x^2(f_{\rm 0}-1)
\end{equation}
with
\begin{equation}
    f_{\rm 0}=\frac{x-\mu}{\phi}
\end{equation}
\begin{equation}
    \phi=\sqrt{1-2\mu x+x^2}
\end{equation}
\begin{equation}
    x=\frac{r_{\rm CN}}{a_{\rm g}}.
\end{equation}
Here $r_{\rm CN}$ is the radius of the condensation nuclei, and $\mu=\cos{\theta_{\rm c}}$ is the cosine of the contact angle of a nucleating substance onto condensation nuclei $\theta_{\rm c}$.
Approximating the embryo surface area as $\pi a_{\rm g}^2$, the heterogeneous nucleation rate, the embryo formation rate per condensation nucleus ($J_{\rm het}~{\rm [s^{-1}]}$), can be evaluated as \citep[e.g.,][]{Moses+92,pruppacher&Klett97}
\begin{equation}\label{eq:J_het}
    J_{\rm het}=4\pi r_{\rm CN}^2\times \pi a_{\rm g}^2\Phi_{\rm v}Zn_{\rm 1,s}\exp{\left(-\frac{\Delta F_{\rm g}f}{k_{\rm B}T}\right)},
\end{equation}
where $n_{\rm 1,s}$ is the concentration of molecules on the nuclei surface.
The concentration on the surface may be determined by the balance between incoming vapor flux $\Phi_{\rm v}$ and outgoing vapor flux $n_{\rm 1,s}\nu_{\rm s}\exp{(-\Delta G_{\rm des}/k_{\rm B}T)}$, i.e.,
\begin{equation}
    n_{\rm 1,s}=\frac{\Phi_{\rm v}}{\nu_{\rm s}}\exp{\left(\frac{\Delta G_{\rm des}}{k_{\rm B}T}\right)},
\end{equation}
where $\nu_{\rm s}$ is the vibration frequency of adsorbed molecules, and $\Delta G_{\rm des}$ is the energy of desorption per molecule.
We note that the number of molecules contained in a critical-sized embryo $g_{\rm *}$ in Equation \eqref{eq:Zeldovich} should be calculated for a spherical cap rather than for a sphere, which is given by
\begin{equation}
    g_{\rm *}=\frac{m_{\rm v}}{\rho_{\rm 0}}\left[\frac{\pi a_{\rm g}^3b^2(3-b)}{3}-\frac{\pi r_{\rm CN}^3c^2(3-c)}{3} \right]
\end{equation}
with
\begin{equation}
   b =1+\left( \frac{1-\mu x}{\phi}\right)
\end{equation}
\begin{equation}
   c =1-f_{\rm 0}.
\end{equation}
Since the shape factor monotonically decreases with an increase in the condensation nuclei size \citep{Frecher58}, large condensation nuclei drastically enhance the heterogeneous nucleation rate.
Moreover, the shape factor also decreases with decreasing the contact angle; thus, a small contact angle is favored to trigger the heterogeneous nucleation \citep[see, e.g., Figure 4 of][]{Lavvas+11a}.

We note that the nucleation rate of Equation \eqref{eq:J_het} is derived under the assumption that vapor molecules are directly deposited to the embryo from the surrounding air.
The surface diffusion of adsorbed molecules to the embryo can also drive the embryo growth, which is generally faster than the nucleation via direct vapor deposition \citep{pruppacher&Klett97}. 
The rate of surface diffusion nucleation can be estimated by \citep{pruppacher&Klett97}
\begin{equation}
    J_{\rm het,diff}\approx J_{\rm het}\exp{\left(\frac{\Delta G_{\rm des}-\Delta G_{\rm sd}}{k_{\rm B}T}\right)},
\end{equation}
where $\Delta G_{\rm sd}$ is the activation energy for the surface diffusion.
Although the activation energy is an uncertain parameter, the energy required to drive the molecular diffusion is likely much smaller than the energy required for the molecular desorption.
Previous studies assumed $\Delta G_{\rm sd}=\Delta G_{\rm des}/10$ \citep{Seki&Hasegawa83,Rannou&West18}.

In addition to the saturation ratio and surface energy, the heterogeneous nucleation rate is a sensitive function of molecular desorption energy $\Delta G_{\rm des}$ and contact angle $\theta_{\rm c}$.
These parameters vary with the combination of the embryo and condensation nucleus substances and are unknown for hydrocarbon ices on photochemical hazes in most cases.
\citet{Rannou+19} derived the desorption energy of $\Delta G_{\rm des}=1.519\times{10}^{-20}~{\rm J}$ and $2.35\times{10}^{-20}~{\rm J}$ for CH$_4$ and C$_2$H$_6$ on the Titan tholin based on a laboratory study of \citet{Curtis+08}.
A few experimental estimations have also been available for contact angles: $\mu=0.994$ ($\theta_{\rm c}=6.27^{\circ}$) for CH$_4$, $\mu=0.966$ ($\theta_{\rm c}=14.98^{\circ}$) for C$_2$H$_6$ \citep{Rannou+19} and C$_4$H$_{10}$ \citep{Curtis+05}.
A recent study of \citet{Yu+20} measured the dispersion and polar components of surface energy for the Titan tholin and used them to predict $\theta_{\rm c}=0^{\circ}$ and $15$--$30^{\circ}$for CH$_4$ and C$_2$H$_6$, respectively.
Based on the same methodology, \citet{Garver+20} suggested the  C$_2$H$_4$ contact angle of $\theta_{\rm c}=0$--${35}^{\circ}$.
Thus, we vary the contact parameter from $\mu=0.99$ to $0.9$ (corresponding to $\theta_{\rm c}=8$--$26^{\circ}$) for C$_2$H$_4$ to check the sensitivity of the heterogeneous nucleation rate on the contact angle.
\citet{Moses+92} provided the surface energy $\sigma$ of various hydrocarbon ices.


}

\end{document}